\documentclass[prd,aps,floats,floatfix,nofootinbib,preprintnumbers]{revtex4-1}
\usepackage{amsmath}
\usepackage{amsfonts}
\usepackage{graphicx}
\usepackage{slashed}
\usepackage{dcolumn}
\usepackage{bm}
\usepackage{amssymb}
\usepackage{latexsym}
\usepackage{color}
\usepackage{subfigure}

\newcommand{\brac}[1]{\left( #1 \right)}

\begin{document}

\preprint{MSU-HEP-111130}
\title{Production of Massive Color-Octet Vector Bosons at Next-to-Leading Order}

\author{R. Sekhar Chivukula}
\email[]{sekhar@msu.edu}
\author{Arsham Farzinnia}
\email[]{farzinni@msu.edu}
\author{Elizabeth H. Simmons}
\email[]{esimmons@msu.edu}
\affiliation{Department of Physics, Michigan State University, East Lansing, MI 48824, USA}

\author{Roshan Foadi}
\email[]{foadiros@msu.edu}
\affiliation{Service de Physique Th\'eorique, Universit\'e Libre de Bruxelles, Brussels, Belgium\\
and Centre for Cosmology, Particle Physics and Phenomenology (CP3)\\
Universit\'e catholique de Louvain, B-1348 Louvain-la-Neuve, Belgium}

\date{December 13, 2011}

\begin{abstract}
We report the first complete calculation of QCD corrections to the production of a massive color-octet vector boson.
Our next-to-leading-order (NLO) calculation includes both virtual corrections as well as corrections arising from the emission of gluons and light quarks, and we demonstrate the reduction in factorization-scale dependence relative to the leading-order
approximation used in previous hadron collider studies. We show that the QCD NLO corrections to coloron production
are as large as 30\%, and that the residual factorization scale-dependence is reduced to of order 2\%. We also calculate the $K$-factor and the $p_T$ spectrum for coloron production, since these are
valuable for comparison with experiment. Our results apply
directly to the production of the massive color-octet vector bosons in axigluon, topcolor, and coloron models,
and approximately to the production of KK gluons in extra-dimensional models or color-octet technivector mesons in
technicolor models.
\end{abstract}

\maketitle

\section{Introduction}

\label{introsec}

Massive color-octet vector bosons are predicted in a variety of models, including axigluon models \cite{Frampton:1987dn,Bagger:1987fz}, topcolor models \cite{Hill:1991at,Hill:1993hs,Popovic:1998vb,Braam:2007pm},
technicolor models with colored technifermions \cite{Chivukula:1995dt}, flavor-universal \cite{Chivukula:1996yr,Simmons:1996fz} and
chiral \cite{Martynov:2009en} coloron models, and extra-dimensional models with KK gluons \cite{Davoudiasl:2000wi,Lillie:2007yh}. These states have also recently been considered
as a potential source  \cite{Ferrario:2009bz,Frampton:2009rk} of the top-quark forward-backward asymmetry observed by the CDF collaboration \cite{Aaltonen:2008hc,Aaltonen:2011kc}.\footnote{Note, however, that the observation of a top-quark forward-backward asymmetry is not confirmed by results of the D0 collaboration \protect\cite{Abazov:2007qb,Abazov:2011rq}. Furthermore, if the observed top-quark forward-backward asymmetry is confirmed, explaining this using color-octet vector bosons is problematic given the tight constraints on flavor-changing neutral-currents \protect\cite{Chivukula:2010fk}.}
Recent searches for resonances in the dijet mass spectrum at the LHC imply that the lower bound on such a boson is now 2-3 TeV \cite{Han:2010rf,Haisch:2011up,CMS,atlas}.\footnote{At least for the fermion charge assignments considered, and in the case where the resonance is narrow compared to the djiet mass resolution of the detector}  If there are color-octet vector bosons associated with the electroweak symmetry breaking sector, as suggested by several of  the models discussed above, their presence should be uncovered by the LHC in the future.

In this paper, we report the first complete calculation\footnote{As this work was being completed, a computation of
the NLO virtual corrections of top-quark pair production via a heavy color-octet vector boson has been reported in \protect\cite{Zhu:2011gd}. That work is complementary to ours in that it does not employ the narrow width approximation for the color-octet boson, but neither does it include real gluon or quark emission. After this work was submitted for publication,
real emission has also been considered by those authors \cite{Zhu:2012um}.} of QCD corrections to the production of a massive color-octet vector boson. We will refer to these massive color-octet vector states generically as ``colorons."  We treat the coloron as an asymptotic state in our calculations, employing the narrow width approximation. Our next-to-leading-order (NLO) calculation includes both virtual corrections as well as corrections arising from the emission of gluons and light quarks, and we demonstrate the reduction in factorization-scale dependence relative to the leading-order (LO)
approximation used in previous hadron collider studies.

The QCD NLO calculation of coloron production reported here differs substantially from the classic computation of the QCD NLO
corrections to Drell-Yan production \cite{Altarelli:1979ub}, because the final state is colored. In particular, Drell-Yan
production involves the coupling of the light quarks to a conserved (or, in the case of $W$- or $Z$-mediated processes,
conserved up to quark masses) current. Hence, in computing the NLO corrections to Drell-Yan processes, the current conservation Ward identity insures a cancellation between the UV divergences arising from virtual
quark wave function and vertex corrections. These cancellations do not occur in the calculation of the NLO corrections to coloron production, because of vertex corrections involving the 3-point non-Abelian colored-boson vertices. As we describe in Section \ref{sec:virual}, we use the ``pinch technique" \cite{Binosi:2009qm} to divide the problematic non-Abelian vertex corrections into two pieces -- a ``pinched" piece whose UV divergence contributes to the renormalization of the coloron wavefunction (and, ultimately, a renormalization of the coloron coupling) and an ``unpinched" part whose UV divergence (when combined with an Abelian vertex correction)
cancels against the UV divergences in quark wavefunction renormalization. As we show, once the UV divergences are properly accounted for, the IR divergences cancel in the usual way: the IR divergences arising from real quark or gluon emission cancel against the IR divergences in the virtual corrections, and the IR divergences arising from collinear quarks or gluons in the initial state are absorbed in the properly defined parton distribution functions (PDFs).

We compute the gauge-, quark-, and self-couplings of the coloron from a theory with
an extended $SU(3)_{1c} \times SU(3)_{2c} \to SU(3)_c$ gauge structure, where $SU(3)_c$ is identified with QCD. The calculation yields
the minimal coupling of gluons to colorons, and
allows for the most general couplings of quarks to colorons.  The cancellation of UV divergences described above, however,
occurs only when the 3-coloron coupling has the strength that arises from the dimension-four gauge-kinetic energy terms of
the extended $SU(3)_{1c} \times SU(3)_{2c}$ gauge structure. Our computation applies directly
to any theory with this structure, {\it i.e.} to massive color-octet vector bosons in axigluon, topcolor, and coloron models.
In general, the triple coupling of KK gluons in extra-dimensional models, or of colored technivector mesons in
technicolor models, will not follow this pattern. However our results apply approximately to these cases
as well, to the extent that the $SU(3)_{1c} \times SU(3)_{2c}$
model is a good low-energy effective theory for the extra dimensional model (a ``two-site" approximation in
the language of deconstruction \cite{ArkaniHamed:2001ca,Hill:2000mu}) or for the technicolor theory (a hidden local symmetry approximation for
the effective technivector meson sector \cite{Bando:1984ej,Bando:1985rf}).\footnote{Arbitary three- and four-point coloron self-couplings
can be incorporated in the $SU(3)_{1c} \times SU(3)_{2c}$ by adding ${\cal O}(p^4)$ terms in the 
of effective chiral Lagrangian of Eq. (\protect\ref{eq:L}), and deviations in these couplings are therefore of ${\cal O}(M^2_C/\Lambda^2)$ where
$\Lambda$ is the cutoff of the effective coloron theory. The 3- and 4-point self-couplings, however, are neither relevant to the leading-order $q\bar{q}$ nor to
the IR divergent NLO coloron production contributions, and therefore numerically insignificant.}

This paper is structured as follows. In Sec.~\ref{formalsec} we introduce the formalism of a minimal vector coloron theory, deriving all the Feynman rules, and setting the stage for the subsequent calculations. In Sec.~\ref{LOsec} we review the leading order computations of the amplitude and cross section for coloron production due to $q\bar{q}$ pair annihilation. Sec.~\ref{sec:virual} describes in detail the one loop virtual corrections to the $q\bar{q}$ pair annihilation process, elaborating on the contributions from the quark self-energy, coloron-coloron and gluon-coloron mixed vacuum polarization amplitudes, and the vertex corrections.
We employ the pinch technique \cite{Binosi:2009qm}, described above, in order to consistently treat the UV divergences, and obtain a gauge-invariant, mutually-independent set of counterterms. The one loop cross section is constructed, and the IR singularities of the virtual correction properly extracted.
In Sec.~\ref{sec:real} we consider the real emission processes, consisting of real (soft and collinear) gluon and (collinear) quark emission.  In Sec.~\ref{sec:NLOcs} we put all the pieces together, exhibiting the explicit cancellation of the IR divergences among the real and virtual corrections, and demonstrate the renormalization of the quark and gluon PDFs. We give a finite expression for the NLO-corrected production cross section.
Finally, in Sec.~\ref{sec:conclusion} we plot the cross section, demonstrate that the QCD NLO corrections are
as large as 30\%, and show that the residual factorization-scale dependence is at the 2\% level.
We also calculate the $K$-factor and the $p_T$ spectrum for coloron production, since these are
valuable for comparison with experiment.

An appendix contains all the Feynman rules of the theory.

\section{A minimal theory for spin-one colorons} \label{formalsec}


In this section, we introduce colorons\footnote{Colorons can in principle be introduced as matter fields in the adjoint of $SU(3)_c$. This approach, however, would lead to an early violation of tree-level unitarity, as the scattering amplitude of longitudinally polarized massive spin-one bosons can grow, by power counting, like $E^4$, where $E$ is the center-of-mass (CM) energy. The only way to avoid this is to ``promote'' the coloron to the status of gauge field of a spontaneously broken gauge theory: then the special relation between trilinear and quartic gauge couplings will lead to an exact cancellation of the terms growing like $E^4$, as happens in the standard electroweak theory.} as the massive color-octet bosons arising when an extended  $SU(3)_{1c}\times SU(3)_{2c}$ gauge symmetry is spontaneously broken by a non-linear sigma model field to its diagonal subgroup, $SU(3)_c$, which we identify with QCD. The symmetry breaking results in a low-energy spectrum that includes both a massless spin-one color octet of gauge bosons, the gluons, and a massive spin-one color octet of gauge bosons, the colorons.

In detail, we replace the QCD Lagrangian with
\begin{eqnarray}
{\cal L}_{\rm color} = - \frac{1}{4}G_{1\mu\nu}G_1^{\mu\nu} - \frac{1}{4}G_{2\mu\nu}G_2^{\mu\nu}
+\frac{f^2}{4}\ {\rm Tr} D_\mu\Sigma \,D^\mu\Sigma^\dagger + {\cal L}_{\rm gauge-fixing} + {\cal L}_{\rm ghost} + {\cal L}_{\rm quark} \ .
\label{eq:L}
\end{eqnarray}
Here $\Sigma$ is the nonlinear sigma field breaking $SU(3)_{1c}\times SU(3)_{2c}$ to $SU(3)_c$,
\begin{eqnarray}
\Sigma = \exp\left(\frac{2 i \pi^a t^a}{f}\right) \ , \quad a=1,\dots,8 \ ,
\end{eqnarray}
where $\pi^a$ are the Nambu-Goldstone bosons ``eaten'' by the coloron, $f$ is the corresponding ``decay-constant", and $t^a$ are the Gell-Mann matrices, normalized as ${\rm Tr}\ t^a t^b=\delta^{ab}/2$. The $\Sigma$ field transforms as the bi-fundamental of $SU(3)_{1c}\times SU(3)_{2c}$,
\begin{eqnarray}
\Sigma\to u_1 \Sigma u_2^\dagger \ , \quad u_i=\exp\left(i \alpha_i^a t^a\right) \ ,
\end{eqnarray}
where the $\alpha_i^a$ are the parameters of the $SU(3)_{ic}$ transformations. This leads to the covariant derivative
\begin{eqnarray}
D_\mu \Sigma = \partial_\mu \Sigma - i g_{s_1} G^a_{1\mu} t^a \Sigma + i g_{s_2} \Sigma\,  G^a_{2\mu} t^a \ ,
\end{eqnarray}
where $g_{s_i}$ is the gauge coupling of the $SU(3)_{ic}$ gauge group. Up to a total divergence, the quadratic terms in the Lagrangian are
\begin{eqnarray}
{\cal L}^{(2)}_{\rm color} &=& \frac{1}{2} G^a_{i\mu}\left(g^{\mu\nu}\partial^2-\partial^\mu\partial^\nu\right)G^a_{i\nu}
+\frac{f^2}{8}\left(g_{s_1} G^a_{1\mu}-g_{s_2} G^a_{2\mu}\right)^2+\frac{1}{2}\left(\partial_\mu\pi^a\right)^2
-\frac{f}{2}\left(g_{s_1} G^a_{1\mu}-g_{s_2} G^a_{2\mu}\right)\partial^\mu\pi^a \nonumber \\
&+& {\cal L}^{(2)}_{\rm gauge-fixing} + {\cal L}^{(2)}_{\rm ghost} + {\cal L}^{(2)}_{\rm quark} \ ,
\end{eqnarray}
where a sum over $i=1,2$ in the gauge kinetic terms is implied.

The gauge-Goldstone mixing term can be removed, up to a total divergence, by choosing the gauge-fixing Lagrangian to be
\begin{eqnarray}
{\cal L}_{\rm gauge-fixing} = -\frac{1}{2} \left({\cal F}_i^a\right)^2~,
\end{eqnarray}
where the gauge-fixing functions are
\begin{eqnarray}
{\cal F}_1^a\equiv \frac{1}{\sqrt{\xi}}\left(\partial^\mu G^a_{1\mu} + \xi \frac{g_{s_1} f}{2}\pi^a\right) \ , \quad
{\cal F}_2^a\equiv \frac{1}{\sqrt{\xi}}\left(\partial^\mu G^a_{2\mu} - \xi \frac{g_{s_2} f}{2}\pi^a\right) \ .
\end{eqnarray}
The Faddeev-Popov ghost Lagrangian is obtained by taking the functional determinant of $\delta {\cal F}_i^a/\delta\alpha_j^b$. This leads to
\begin{eqnarray}
{\cal L}_{\rm ghost} = \bar{c}_i^a \Big[-\partial^\mu\left(\delta_{ij}\delta^{ab}\partial_\mu
-g_{s_i} f^{abc} \delta_{ij} G_{i\mu}^c\right) -\xi \frac{g_{s_i}^2 f^2}{4}\left(\delta_{i1}-\delta_{i2}\right)\left(\delta_{1j}-\delta_{2j}\right)
\delta^{ab}+{\cal O}(\pi)\Big] c_j^b \ ,
\end{eqnarray}
where $f^{abc}$ are the $SU(3)$ structure constants, and a sum over $i,j=1,2$ is implied. Notice that we have included only the inhomogeneous terms in the transformation of the ``eaten'' Goldstone boson, whence the unspecified ${\cal O}(\pi)$ terms in the ghost Lagrangian, which are unnecessary for our computation. Up to a total divergence, the quadratic Lagrangian now reads
\begin{eqnarray}
{\cal L}^{(2)}_{\rm color} &=& \frac{1}{2} G^a_{i\mu}\Bigg[\delta_{ij} g^{\mu\nu} \partial^2
-\delta{ij}\left(1-\frac{1}{\xi}\right)\partial^\mu\partial^\nu + \frac{g_{s_i}^2 f^2}{4}
\left(\delta_{i1}-\delta_{i2}\right)\left(\delta_{1j}-\delta_{2j}\right)\Bigg] G^a_{j\nu}
- \frac{1}{2}\pi^a\Bigg[\partial^2+\frac{\xi}{4}\left(g_{s_1}^2+g_{s_2}^2\right)f^2\Bigg]\pi^a \nonumber \\
&-& \bar{c}_i^a \Bigg[\delta_{ij}\partial^2 + \xi\frac{g_{s_i}^2 f^2}{4}
\left(\delta_{i1}-\delta_{i2}\right)\left(\delta_{1j}-\delta_{2j}\right)\Bigg] c_j^a + {\cal L}^{(2)}_{\rm quark} \ .
\label{eq:quadratic}
\end{eqnarray}

Aside from a factor of the gauge-fixing parameter $\xi$, the gauge and ghost fields share the same mass matrix, as expected. This is diagonalized by
\begin{eqnarray}
\left(\begin{array}{c} G^a_{1\mu} \\ G^a_{2\mu} \end{array}\right) = R\ \left(\begin{array}{c} G^a_\mu \\ C^a_\mu \end{array}\right) \ ,
\quad
\left(\begin{array}{c} c^a_1 \\ c^a_2 \end{array}\right) = R\ \left(\begin{array}{c} c_G^a \\ c_C^a \end{array}\right) \ ,
\label{eq:rotations}
\end{eqnarray}
where
\begin{eqnarray}
R\equiv \left(\begin{array}{cc}\cos\theta_c & -\sin\theta_c \\ \sin\theta_c & \cos\theta_c \end{array}\right) \ , \quad
\sin\theta_c \equiv \frac{g_{s_1}}{\sqrt{g_{s_1}^2+g_{s_2}^2}} \ \ .
\end{eqnarray}
In Eq.~(\ref{eq:rotations}) $G^a_\mu$ is the gluon field and $C^a_\mu$ is the coloron field, whereas $c_G^a$ and $c_C^a$ are the corresponding ghost fields. Inserting these expressions in Eq.~(\ref{eq:quadratic}) gives, for the coloron mass,
\begin{eqnarray}
M_C = \frac{\sqrt{g_{s_1}^2+g_{s_2}^2}\ f}{2}\equiv \frac{g_s\ f}{\sin 2\theta_c} \ ,
\end{eqnarray}
where $g_s$ is the $SU(3)_c$ coupling,
\begin{eqnarray}
\frac{1}{g_s^2} = \frac{1}{g_{s_1}^2}+\frac{1}{g_{s_2}^2}\ .
\end{eqnarray}
The gluon ghost is massless, whereas both the coloron ghost and the eaten Goldstone boson have mass $\sqrt{\xi}M_C$. The interaction vertices and the corresponding Feynman rules can be found in Appendix~\ref{sec:FRs}.

We will leave the quark charge assignments under $SU(3)_{1c}\times SU(3)_{2c}$ arbitrary, for greater generality. In the mass eigenstate basis we write
\begin{equation} \label{Lferm}
\mathcal{L}_{\rm quark} = \bar{q}^i i\left[\slashed{\partial}-i g_s \slashed{G}^a t^a -i \slashed{C}^a t^a \left(g_L P_L+g_R P_R\right)\right]q_i~,
\end{equation}
where $P_L$ and $P_R$ are the helicity projection operators,
\begin{eqnarray}
P_L\equiv \frac{1-\gamma_5}{2}\ , \quad  P_R\equiv \frac{1+\gamma_5}{2} \ ,
\end{eqnarray}
and $i$ is a flavor index\footnote{Here we work in the broken electroweak phase, and only employ fermion mass eigenstates.}. The coupling to the gluon is dictated by charge universality, whereas the $g_L$ and $g_R$ couplings to the coloron depend on the original charge assignments of the quarks. For example, if both left-handed and right-handed quarks are only charged under $SU(3)_{1c}$, then $g_L=g_R=-g_s\tan\theta_c$, while the axigluon \cite{Frampton:1987dn,Bagger:1987fz} corresponds to
$g_L=-g_R=g_s$ ({\it i.e.} $\theta_c=\pi/4$). In general, $g_L$ and $g_R$ can each take on the values $-g_s\tan\theta_c$ or  $g_s\cot\theta_c$ in any specific model,\footnote{It is possible to generalize this setup to non-universal charge assignments: in this case flavor-diagonal chiral couplings to the coloron would depend on a generation index. Flavor-changing couplings are strongly constrained \protect\cite{Chivukula:2010fk}.}
\begin{equation} \label{gLgR}
g_L, g_R \in \left \{ -g_s \tan \theta_c, g_s \cot \theta_c \right \} \ .
\end{equation}
\section{LO Coloron Production} \label{LOsec}
\begin{figure}[!t]
\begin{center}
\includegraphics[width=1.5in]{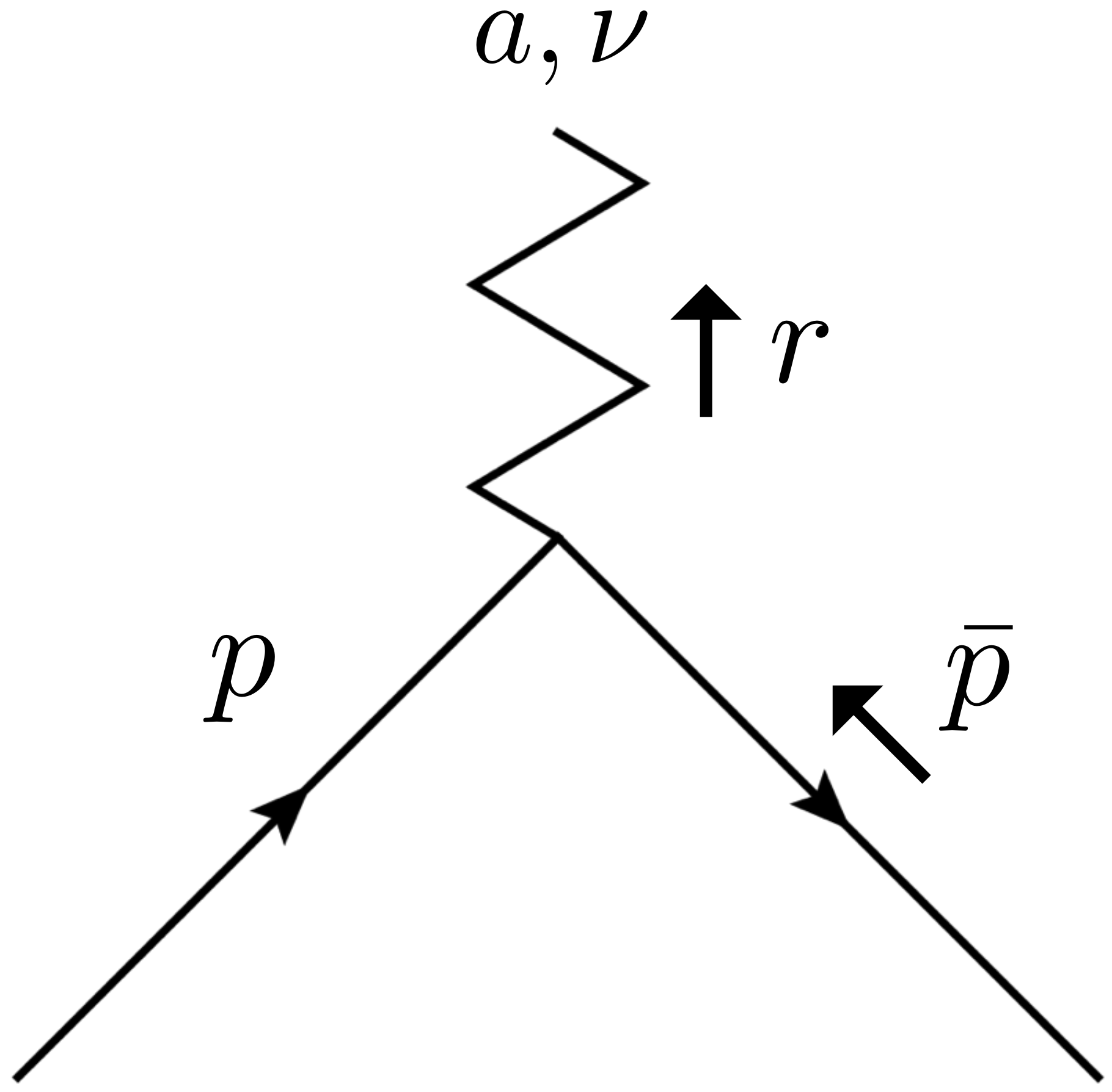}
\caption{Tree-level contribution to coloron production.  The coloron field, $C^a_\nu$, is represented by the zigzag line.}
\label{LOfig}
\end{center}
\end{figure}
The dominant channel for coloron production at a hadron collider is given by the tree-level diagram of Fig.~\ref{LOfig}, in which a $q\bar{q}$ pair annihilates into a coloron. The tree-level diagram with gluon-gluon fusion into a coloron does not exist in the Lagrangian of Eq.~(\ref{eq:L}): in general there are no dimension-four terms with two gauge bosons of an unbroken symmetry and a spin-one field charged under the same symmetry. We use the narrow width approximation for the coloron, take the quarks to be on-shell, and set their masses to zero: this is certainly a good approximation, as current experimental bounds
\cite{Han:2010rf,CMS,atlas}  constrain the coloron mass to be in the TeV range.

The leading order (LO) amplitude corresponding to the diagram of Fig.~\ref{LOfig} is
\begin{equation}
i \mathcal{M}_{q\bar{q}\to C}^{(0)}
= g_s\ \bar{v}^r(\bar{p}) \ i \gamma^{\mu} \left(r_L P_L+r_R P_R \right) t^a \ u^s (p) \ \varepsilon_\mu^{a\lambda\ast}(r)~,
\label{eq:tree}
\end{equation}
where the superscripts $r$ and $s$ denote quark spin projections, $\lambda$ is the coloron polarization, and
\begin{equation}
r_L\equiv \frac{g_L}{g_s}\ , \quad r_R\equiv \frac{g_R}{g_s} \ , \quad r_L, r_R \in \left \{-\tan \theta_c, \cot \theta_c \right \} \ .
\end{equation}
In $d=2(2-\epsilon)$ dimensions the squared amplitude averaged over initial spins and colors, and summed over final polarization states, is
\begin{align}
\overline{|\mathcal{M}_{q\bar{q}\to C}^{(0)}|^2} \equiv  \left(\frac{1}{{\rm dim}(r)}\right)^{2} \left(\frac{1}{2}\right)^2
\sum_{\text{spin \& color}} |\mathcal{M}_{q\bar{q}\to C}^{(0)}|^2
= \frac{C_2(r)(1-\epsilon)}{2\ {\rm dim}(r)}\ g_s^2 \left(r_L^2+r_R^2 \right) \hat{s}
\label{m0sqav} \ ,
\end{align}
where ${\rm dim}(r)=3$ and $C_2(r)=4/3$ are respectively the dimension and Casimir of the fundamental representation of $SU(3)$, and $\hat{s}\equiv (p+\bar{p})^2=2 p \cdot \bar{p}$ is the partonic center-of-mass CM energy. This gives the LO cross section \cite{Bagger:1987fz} for $q\bar{q}\to C$,  
\begin{align}
\hat{\sigma}_{q\bar{q}\to C}^{(0)} = \frac{\pi}{\hat{s}^2} \ \overline{|\mathcal{M}_{q\bar{q}\to C}^{(0)}|^2} \ \delta(1-\chi)
=\frac{\alpha_s A (r_L^2+r_R^2)}{\hat{s}}\ \delta(1-\chi)\ ,
\label{cstree}
\end{align}
where $\alpha_s\equiv g_s^2/4\pi$,
\begin{equation}
A \equiv \frac{2\pi^2 C_2(r)(1-\epsilon)}{{\rm dim}(r)} \ ,
\label{eq:defA}
\end{equation}
and
\begin{equation}
\chi \equiv \frac{M_C^2}{\hat{s}} \ .
\label{eq:chi}
\end{equation}

The full LO cross section for $pp\to C$ is given by the convolution of the LO partonic cross section $\hat{\sigma}_{q\bar{q}\to C}^{(0)}$ with the parton distribution functions (PDFs) for the quarks within the protons, and a sum over all quark flavors,
\begin{equation}
\sigma^{LO} = \int dx_1\int dx_2 \sum_q \Big[ f_q(x_1) f_{\bar{q}}(x_2) + f_{\bar{q}}(x_1) f_q(x_2) \Big]
\hat{\sigma}_{q\bar{q}\to C}^{(0)} \ ,
\label{CSfulldetect}
\end{equation}
where $f_q(x)$ is the PDF of parton $q$, and $x$ the momentum fraction of the corresponding parton. Taking the collision axis to be the 3-axis, the four-momenta
of the partons are:
\begin{eqnarray}
p=\frac{\sqrt{s}}{2}\left(x_1,0,0,x_1\right)\ ,\quad
\bar{p}=\frac{\sqrt{s}}{2}\left(x_2,0,0,-x_2\right) \ ,
\end{eqnarray}
where $s$ is the CM energy of the colliding hadrons. This gives
\begin{eqnarray}
\hat{s}=x_1\ x_2\ s \ , \quad
\chi = \frac{M_C^2}{s\ x_1\ x_2} \ .
\label{eq:x1x2}
\end{eqnarray}
\section{NLO Coloron Production: Virtual Corrections} \label{sec:virual}

In this section we compute the next-to-leading order (NLO) virtual QCD corrections to the $q\bar{q}\to C$ amplitude. These include one-loop wave-function and vertex corrections, which we choose to compute in 't Hooft-Feynman gauge, $\xi=1$. The non-Abelian vertex corrections are computed by employing the {\it pinch technique}: this allows us to obtain QED-like Ward identities, and absorb all UV infinities in the renormalization of the gauge field propagators. After inclusion of the counterterms, the virtual corrections are UV-finite, yet IR infinite. In Sec.~\ref{sec:real} we show that the IR divergences cancel once the real corrections, corresponding to the emission of soft  and collinear gluons and quarks, are included in the calculation of the inclusive production cross section. Our loop integrals are computed in dimensional regularization, with $d=2(2-\epsilon)$ dimensions. We first regulate the IR divergences by giving the gluon a small mass ($m_g \to 0^+$): in this way all infinities are in the UV, and regularization requires $\epsilon>0$. After all of the UV infinities are removed, by cancellation and inclusion of the counterterms, we let the gluon mass approach zero. This will make the virtual corrections IR divergent, with the infinities being regulated by taking $\epsilon<0$.

\begin{figure}[!t]
\begin{center}
\includegraphics[width=4.5in]{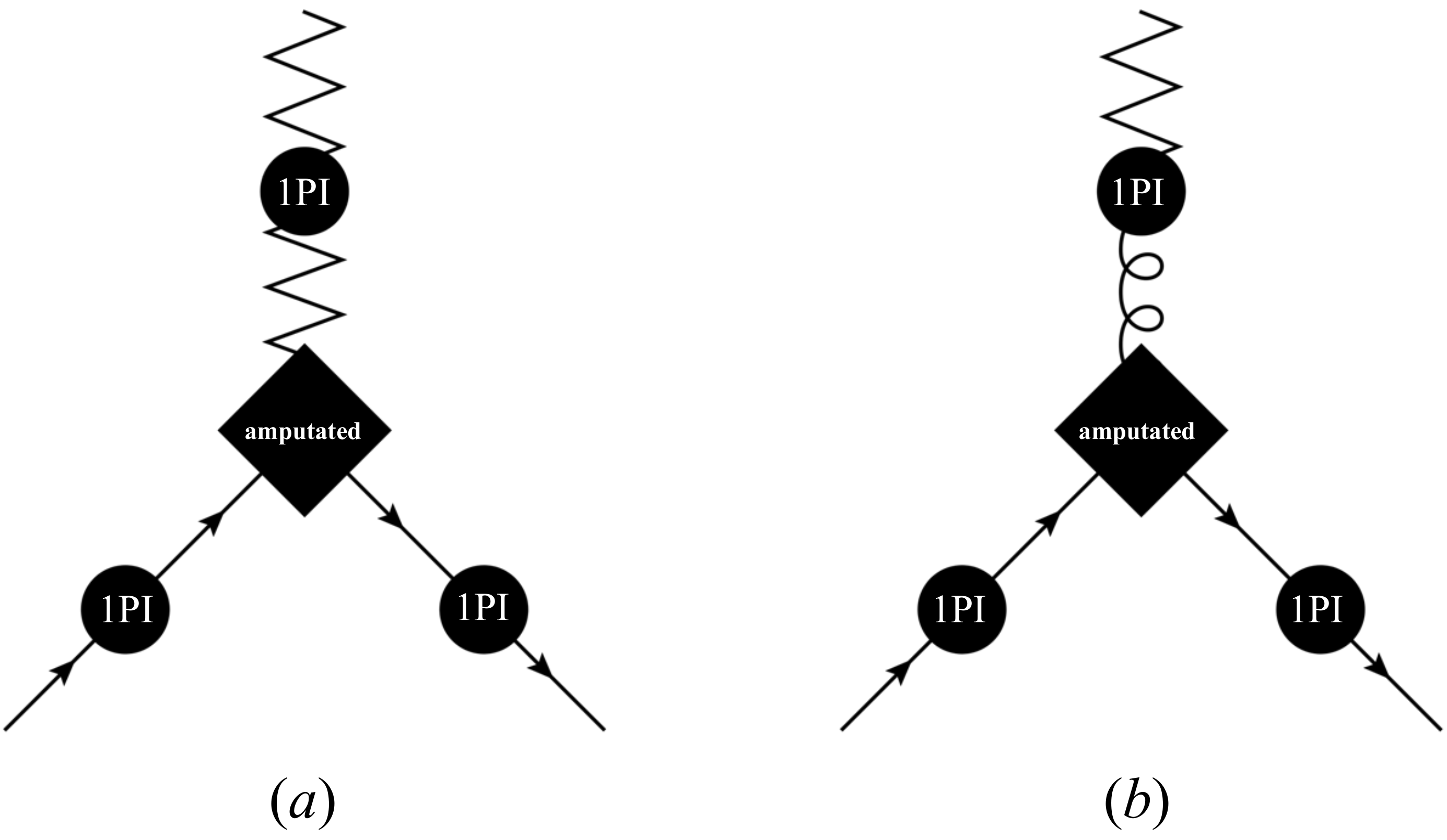}
\caption{Structure of $q\bar{q} \to C$ amplitude, to all orders in perturbation theory. Direct coloron production is illustrated on the left, while production via mixing
with the gluon is shown on the right.  The gluon field is, as usual, represented by the coiling line; the coloron field is represented by the zigzag line.}
\label{fig:lsz}
\end{center}
\end{figure}

\begin{figure}[!t]
\begin{center}
\includegraphics[width=4.5in]{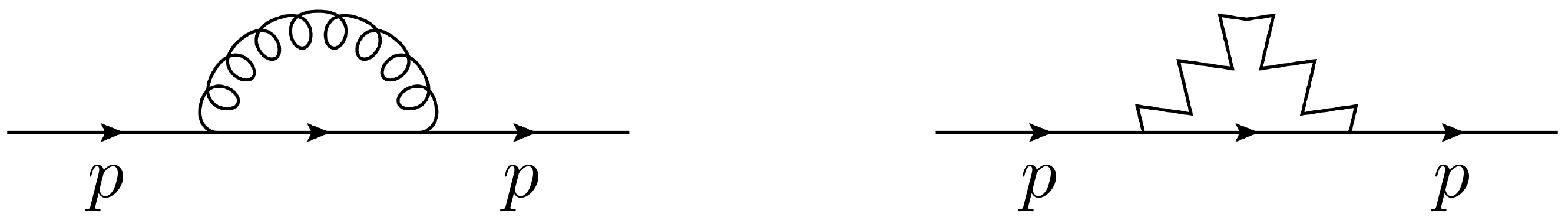}
\caption{Quark self-energy diagrams at one loop.  Particle notation as defined in Fig. \protect\ref{fig:lsz}.}
\label{fig:self}
\end{center}
\end{figure}

Since the quark couplings to the coloron are chiral, in general, we need a prescription for treating $\gamma_5$ in $d\neq 4$. Here we take $\gamma_5$ to always anticommute with $\gamma^\mu$. Choosing an alternative prescription, such as 't Hooft-Veltman in which $\gamma_5$ anticommutes with $\gamma^\mu$ for $\mu=0,1,2,3$ and commutes for other values of $\mu$, would lead to a cross section for $q\bar{q}\to C$ which differs from ours by only a finite renormalization of the coupling(s). 


The general structure of the $q\bar{q}\to C$ amplitude, illustrated in Fig. \ref{fig:lsz},  is
\begin{eqnarray}
i\mathcal{M}_{q\bar{q}\to C} = g_s\ \bar{v}^r(\bar{p})\ i\left[Z_C^{1/2}\Gamma_{qqC}^{a\mu}+\Gamma^{a\mu}_{qqG} \frac{\Pi_{GC}(\hat{s})}{\hat{s}} \right]
Z_q u^s (p) \ \varepsilon_\mu^{a\lambda\ast}(r) \ ,
\label{eq:M}
\end{eqnarray}
where $\Gamma^{a\mu}_{qqC}$ ($\Gamma^{a\mu}_{qqG}$) is the one-particle-irreducible (1PI) quark-quark-coloron (quark-quark-gluon) vertex and $\Pi_{GC}$ is the coefficient of $g^{\mu\nu}$ in the gluon-coloron vacuum polarization mixing amplitude (VPA).  The factors $Z_q$ and $Z_C$ are, respectively, the residues of the full quark and coloron propagators at the mass pole; they are obtained from the quark self-energy amplitude, $\Sigma(\slashed{p})$, and the coefficient of $g^{\mu\nu}$  in the coloron-coloron VPA, $\Pi_{CC}(q^2)$, as follows:
\begin{eqnarray}
Z_q=\frac{1}{1-\Sigma^\prime(0)} \ , \quad
Z_C=\frac{1}{1-\Pi^\prime_{CC}(M^2_{C {\rm phys}})} \ ,
\label{eq:Z}
\end{eqnarray}
where the prime denotes a derivative with respect to the argument, and $M_{C {\rm phys}}$ is the coloron's physical mass. To lowest order, $Z_q=1$, $Z_C=1$, $\Pi_{GC}=0$, and $i\Gamma^{a\mu}_{qqC}=\gamma^\mu \left(r_L P_L+r_R P_R \right) t^a$; inserting these in Eq. (\ref{eq:M}) recovers the tree-level amplitude of Eq.~(\ref{eq:tree}).
\subsection{Quark Self-Energy} \label{sec:self}
The NLO quark self-energy correction to the $q\bar{q}\to C$ amplitude is found, from Eqs.~(\ref{eq:M}) and (\ref{eq:Z}), to be
\begin{equation}
i Q = \bar{v}^r(\bar{p}) \ i \gamma^{\nu} \left(g_L P_L+g_R P_R \right) t^a\ \delta Z_q \ u^s (p) \ \varepsilon_\nu^{a\lambda\ast}(r)~,
\label{eq:Q1}
\end{equation}
where
\begin{eqnarray}
\delta Z_q = \Sigma^\prime(0) \ .
\label{eq:Zq}
\end{eqnarray}
At one loop, the $\Sigma(\slashed{p})$ amplitude is given by the diagrams of Fig.~\ref{fig:self}. These lead to the expression
\begin{eqnarray}
\Sigma(\slashed{p})=-\slashed{p}\ \frac{g_s^2 C_2(r)2(1-\epsilon)\Gamma(\epsilon)}{16\pi^2} \int_0^1 dx\ (1-x)
\left[\left(\frac{4\pi\mu^2}{\Delta_{Gq}}\right)^\epsilon  +
\left(\frac{4\pi\mu^2}{\Delta_{Cq}}\right)^\epsilon (r_L^2 P_L+r_R^2 P_R)\right] \ ,
\label{eq:Sigma}
\end{eqnarray}
where $\Gamma(\epsilon)$ is the Euler Gamma-function evaluated at infinitesimal $\epsilon$, and
\begin{eqnarray}
\Delta_{Gq}=(1-x)m_g^2-x(1-x)p^2-i\eta \ , \quad
\Delta_{Cq}=(1-x)M_C^2-x(1-x)p^2-i\eta \ .
\end{eqnarray}
The parameter $\mu$ is the mass scale introduced by the loop integral in $d$ dimensions, and $\eta$ is the positive infinitesimal parameter giving the appropriate prescription for computing the integral in momentum space. As previously anticipated, we have introduced a small gluon mass, $m_g$, in order to regulate the IR divergences and isolate the UV infinities: with $m_g\neq 0$, $\Sigma(\slashed{p})$ and $\Sigma^\prime(\slashed{p})$ contain only UV divergences. Inserting Eq.~(\ref{eq:Sigma}) in Eq.~(\ref{eq:Zq}) gives
\begin{eqnarray}
\delta Z_q = -\ \frac{g_s^2 C_2(r)2(1-\epsilon)\Gamma(\epsilon)}{16\pi^2} \int_0^1 dx\ (1-x)
\left[\left(\frac{4\pi\mu^2}{(1-x)m_g^2-i\eta}\right)^\epsilon  +
\left(\frac{4\pi\mu^2}{(1-x)M_C^2-i\eta}\right)^\epsilon (r_L^2 P_L+r_R^2 P_R)\right]~.
\label{eq:Qp}
\end{eqnarray}
The amplitude of Eq.~(\ref{eq:Q1}) becomes
\begin{eqnarray}
i Q = -\ \frac{\alpha_s}{4\pi}\ 2 C_2(r)(1-\epsilon)\Gamma(\epsilon) \int_0^1 dx \int_0^{1-x} dy
\left[\left(\frac{4\pi\mu^2}{(1-x)m_g^2-i\eta}\right)^\epsilon i \mathcal{M}_{q\bar{q}\to C}^{(0)} +
\left(\frac{4\pi\mu^2}{(1-x)M_C^2-i\eta}\right)^\epsilon i \mathcal{M}_{q\bar{q}\to C}^{\prime(0)}\right]~,
\label{eq:Q}
\end{eqnarray}
where $\mathcal{M}_{q\bar{q}\to C}^{(0)}$ is given by Eq.~(\ref{eq:tree}), and
\begin{equation}
i \mathcal{M}_{q\bar{q}\to C}^{\prime(0)} = g_s\ \bar{v}^r(\bar{p}) \ i \gamma^{\nu} \left(r_L^3 P_L+r_R^3 P_R \right) t^a \ u^s (p) \
\varepsilon_\nu^{a\lambda\ast}(r)  \ .
\label{eq:selfloop}
\end{equation}
For later convenience we have traded the $1-x$ factor, in Eq.~(\ref{eq:Qp}), for an integral over $dy$: this will allow us to directly add the self-energy correction to the vertex correction and explicitly show the cancellation of the UV divergences.
\subsection{Abelian Vertex Corrections} \label{sec:Abelian}
\begin{figure}[!t]
\includegraphics[width=3.0in]{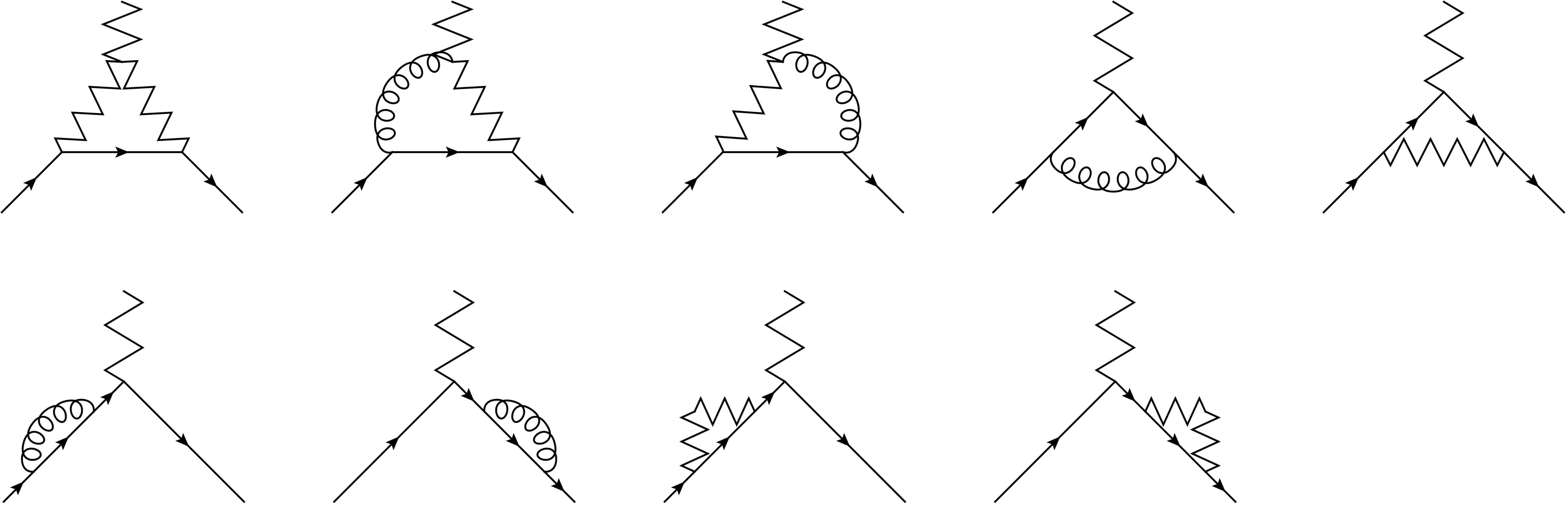}
\caption{One-loop Abelian vertex correction to the $q\bar{q}\to C$ amplitude.   Particle notation is as defined in Fig. \protect\ref{fig:lsz}.}
\label{fig:AV}
\end{figure}
The one-loop Abelian vertex correction to the $q\bar{q}\to C$ amplitude is given by the diagrams of Fig.~\ref{fig:AV}. These lead to the amplitude
\begin{eqnarray}
i V_{\rm Abelian} &=& \frac{\alpha_s}{4\pi} \left[2C_2(r)-C_2(G)\right] \Gamma(1+\epsilon)\int_0^1 dx \int_0^{1-x} dy \Bigg\{\Bigg[
\frac{(1-\epsilon)^2}{\epsilon}-\left(x y \epsilon-(1-x)(1-y)\right)\frac{\hat{s}}{\Delta_{Gqq}}\Bigg]
\left(\frac{4\pi\mu^2}{\Delta_{Gqq}}\right)^\epsilon i\mathcal{M}_{q\bar{q}\to C}^{(0)} \nonumber \\
&+& \Bigg[
\frac{(1-\epsilon)^2}{\epsilon}-\left(x y \epsilon-(1-x)(1-y)\right)\frac{\hat{s}}{\Delta_{Cqq}}\Bigg]
\left(\frac{4\pi\mu^2}{\Delta_{Cqq}}\right)^\epsilon i\mathcal{M}_{q\bar{q}\to C}^{\prime(0)} \Bigg\} \ ,
\label{eq:VA}
\end{eqnarray}
where $C_2(G)=3$ is the Casimir of the adjoint representation, and
\begin{eqnarray}
\Delta_{Gqq} &=& (1-x-y)m_g^2 - x y \hat{s}- i \eta~, \nonumber \\
\Delta_{Cqq} &=& (1-x-y)M_C^2 - x y \hat{s}- i \eta \ .
\end{eqnarray}
Once again, we have included a small gluon mass $m_g$ in order to regulate the IR divergences.
\subsection{Non-Abelian Vertex Corrections {\em a la} Pinch-Technique: Unpinched Diagrams} \label{sec:nonpinched}
\begin{figure}[!t]
\includegraphics[width=5.0in]{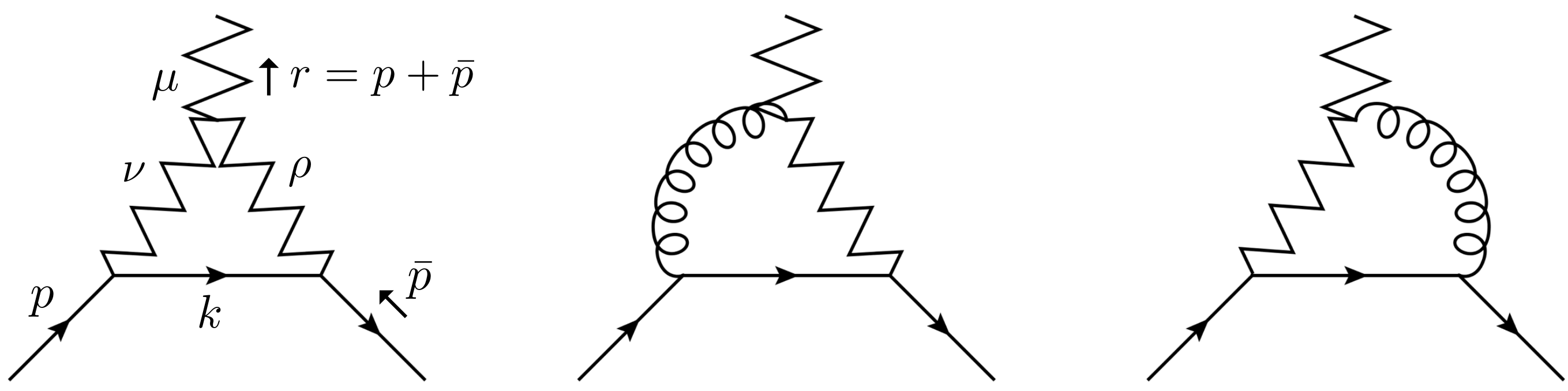}
\caption{One-loop non-Abelian vertex correction to the $q\bar{q}\to C$ amplitude.  Particle notation is as defined in Fig. \protect\ref{fig:lsz}.  Each three-gauge-boson vertex in these diagrams is a full non-Abelian vertex $\Gamma^{\mu\nu\rho}$ in Eq. (\ref{eq:refa}).}
\label{fig:NAV}
\end{figure}

The non-Abelian vertex corrections are given by the diagrams of Fig.~\ref{fig:NAV}. When added to the overall Abelian vertex correction, Eq.~(\ref{eq:VA}), these give the one-loop total vertex correction to $q\bar{q}\to C$. Unlike in QED, the UV divergences in the vertex correction do not cancel the UV divergences arising from the self-energy amplitudes. The reason for this is that the QED Ward identity $\partial^\mu j_\mu=0$ is now replaced by its non-Abelian counterpart $D^\mu j^a_\mu=0$, which does not imply the equality of vertex and quark-wavefunction renormalization constants. It is possible, though, to recover QED-like Ward identities for the currents $j^a_\mu$ by employing the pinch technique. This consists of breaking up the gauge-boson internal momenta of a Feynman diagram into ``pinching'' and ``non-pinching'' pieces. The pinching momenta are those which cancel some internal propagators, leading to a simpler diagram with the external-momentum structure of a propagator. The non-pinching momenta will instead give overall amplitudes satisfying QED-like Ward identities. A formal proof of these statements, for an arbitrary non-Abelian gauge theory, can be found in the review of Ref.~\cite{Binosi:2009qm} (and references therein).

In our vertex computation the pinch technique works as follows. The non-Abelian vertex structure in each of the diagrams in Fig.~\ref{fig:NAV} is
\begin{eqnarray}
\Gamma^{\mu\nu\rho}(k,p,\bar{p})=g^{\mu\nu}(-2p-\bar{p}+k)^\rho + g^{\nu\rho}(p-\bar{p}-2k)^\mu
+g^{\rho\mu}(k+p+2\bar{p})^\nu \ .
\label{eq:refa}
\end{eqnarray}
We can break this into two parts,
\begin{eqnarray}
\Gamma^{\mu\nu\rho}(k,p,\bar{p})=\Gamma_F^{\mu\nu\rho}(k,p,\bar{p}) + \Gamma_P^{\mu\nu\rho}(k,p,\bar{p}) \ ,
\label{eq:breakup}
\end{eqnarray}
where
\begin{eqnarray}
\Gamma_F^{\mu\nu\rho}(k,p,\bar{p})&=&-2 g^{\mu\nu}(p+\bar{p})^\rho + 2 g^{\rho\mu}(p+\bar{p})^\nu
+ g^{\nu\rho}(p-\bar{p}-2k)^\mu \ , \label{eq:RR}\\
\Gamma_P^{\mu\nu\rho}(k,p,\bar{p})&=& g^{\mu\nu}(\bar{p}+k)^\rho + g^{\rho\mu}(k-p)^\nu \ .
\label{eq:breakup2}
\end{eqnarray}
Unlike $\Gamma^{\mu\nu\rho}(k,p,\bar{p})$, the $\Gamma_F^{\mu\nu\rho}(k,p,\bar{p})$ vertex satisfies a QED-like Ward identity for the $g C\to C$ and $C C\to C$ amplitudes,
\begin{eqnarray}
(p+\bar{p})_\mu \Gamma_F^{\mu\nu\rho}(k,p,\bar{p}) = g^{\nu\rho} \Big[(p-k)^2-(\bar{p}+k)^2\Big] \ .
\label{eq:qedWI}
\end{eqnarray}
As shown below, when $\Gamma_F^{\mu\nu\rho}(k,p,\bar{p})$  is used to compute the integral in momentum space (instead of  $\Gamma^{\mu\nu\rho}(k,p,\bar{p})$), its UV divergences, added to the UV divergences of the Abelian vertex corrections, exactly cancel the UV divergences of the quark self-energy amplitudes. As mentioned above, this occurs because a QED-like Ward identity for $q\bar{q}\to C$ holds, as one can prove by using the QED-like Ward identity for the $g C\to C$ and $C C\to C$ amplitudes given in Eq.~(\ref{eq:qedWI}). The three diagrams which correspond to using $\Gamma_F^{\mu\nu\rho}(k,p,\bar{p})$ instead of $\Gamma^{\mu\nu\rho}(k,p,\bar{p})$ are symbolically denoted with a black disk over the non-Abelian vertex, and are shown in Fig.~\ref{fig:NAVF}. These lead to the following contribution to the $q\bar{q}\to C$ amplitude:
\begin{figure}[!t]
\includegraphics[width=5.0in]{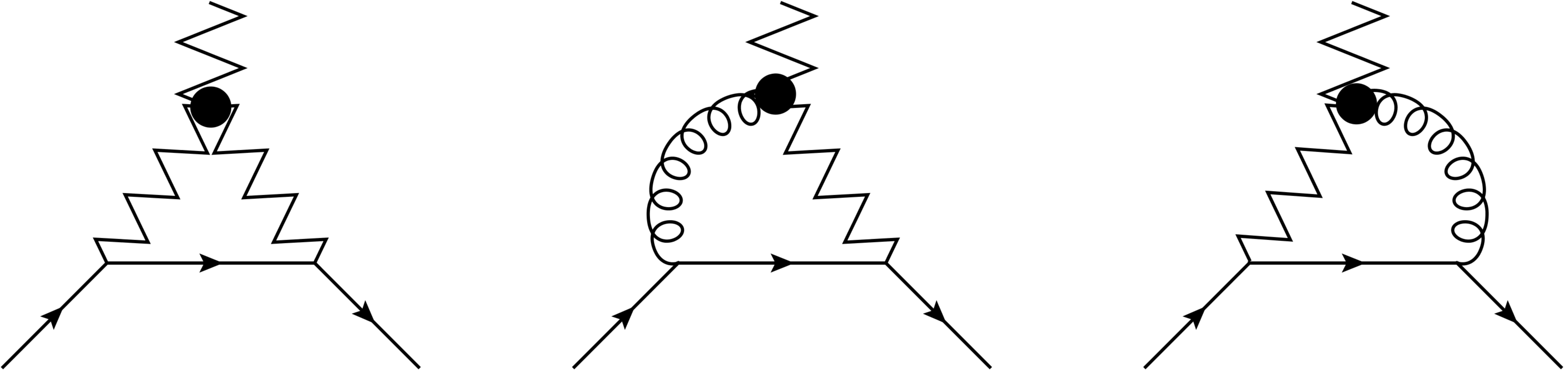}
\caption{Non-Abelian un-pinched vertex-correction diagrams for the $q\bar{q}\to C$ amplitude at one loop.  Particle notation is as defined in Fig. \protect\ref{fig:lsz}. The black disk indicates that each three-point gauge boson vertex in these diagrams has been replaced by the non-pinched portion $\Gamma_F^{\mu\nu\rho}$, as described in Eqs. (\protect\ref{eq:breakup}) and (\protect\ref{eq:RR}). }
\label{fig:NAVF}
\end{figure}
\begin{eqnarray}
i V_{\rm non-Abelian} &=& \frac{\alpha_s}{4\pi} C_2(G)\Gamma(1+\epsilon) \int_0^1 dx \int_0^{1-x} dy \Bigg\{
\Bigg[\left(\frac{1-\epsilon}{\epsilon}-(x+y)\frac{\hat{s}}{\Delta_{GCq}}\right)\left(\frac{4\pi\mu^2}{\Delta_{GCq}}\right)^\epsilon \nonumber \\
&+&\left(\frac{1-\epsilon}{\epsilon}- (x+y)\frac{\hat{s}}{\Delta_{CGq}}\right)\left(\frac{4\pi\mu^2}{\Delta_{CGq}}\right)^\epsilon
-\left(\frac{1-\epsilon}{\epsilon}-(x+y)\frac{\hat{s}}{\Delta_{CCq}}\right)\left(\frac{4\pi\mu^2}{\Delta_{CCq}}\right)^\epsilon
\Bigg] i \mathcal{M}_{q\bar{q}\to C}^{(0)} \nonumber \\
&+&\left(\frac{1-\epsilon}{\epsilon}-(x+y)\frac{\hat{s}}{\Delta_{CCq}}\right)\left(\frac{4\pi\mu^2}{\Delta_{CCq}}\right)^\epsilon
i \mathcal{M}_{q\bar{q}\to C}^{\prime(0)} \Bigg\} \ ,
\label{eq:NAV}
\end{eqnarray}
where
\begin{eqnarray}
\Delta_{GCq} &=& x m_g^2 + y M_C^2 - x y \hat{s} - i\eta~, \nonumber \\
\Delta_{CGq} &=& x M_C^2 + y m_g^2 - x y \hat{s} - i\eta~, \nonumber \\
\Delta_{CCq} &=& (x+y)M_C^2 - x y \hat{s} - i\eta \ .
\end{eqnarray}
In order to obtain Eq.~(\ref{eq:NAV}) we have used the equations of motion for the external spinors, together with the relations
\begin{eqnarray}
2 \cot 2\theta\ r_L = -1 + r_L^2 \ , \quad
2 \cot 2\theta\ r_R = -1 + r_R^2 \ ,
\label{eq:couplingrel}
\end{eqnarray}
which are true for any charge assignment of the quarks. As anticipated, $i Q+i V_{\rm Abelian}+ i V_{\rm non-Abelian}$ is free of UV divergences, as manifestly shown by adding together Eqs.~(\ref{eq:Q}), (\ref{eq:VA}) and (\ref{eq:NAV}). This part of the amplitude is however IR divergent in the limit of zero gluon mass. Setting $m_g= 0$ and $\epsilon<0$ gives
\begin{eqnarray}
i Q+i V_{\rm Abelian}+ i V_{\rm non-Abelian} = \frac{\alpha_s}{4\pi}\Bigg[ C_2(r)\left(-\frac{2}{\epsilon^2}
-\frac{3+2i}{\epsilon}\right)
+ C_2(G) \frac{i\pi}{\epsilon} \Bigg] i \mathcal{M}_{q\bar{q}\to C}^{(0)} + {\rm finite} \ .
\end{eqnarray}

\begin{figure}[b]
\includegraphics[width=4.5in]{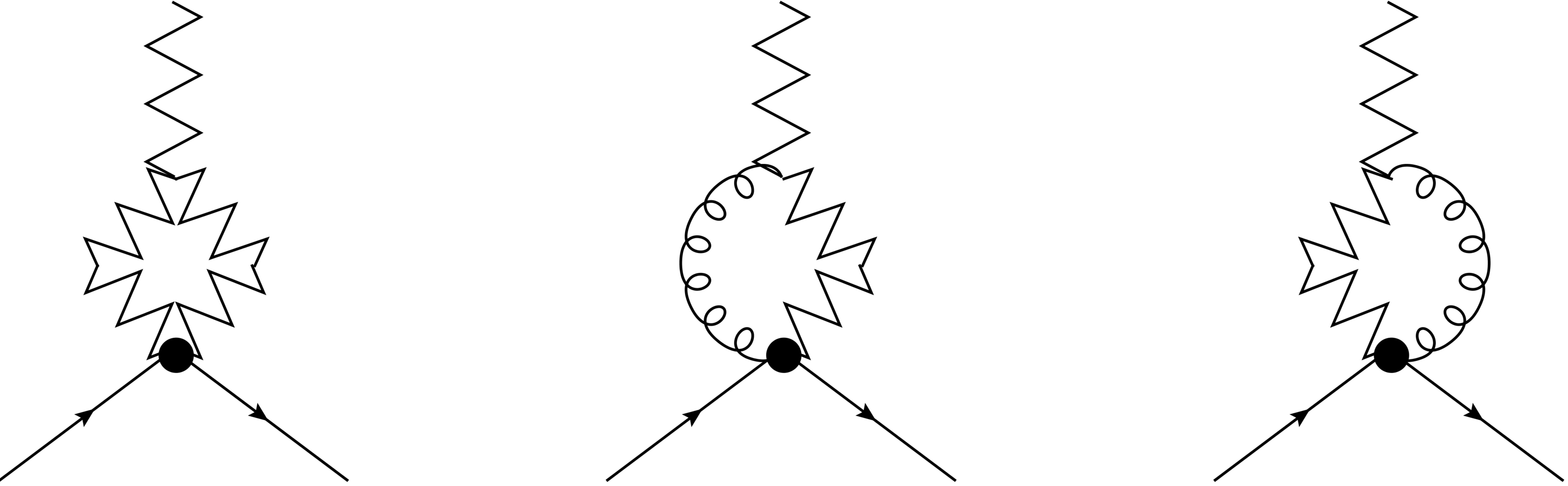}
\caption{Non-Abelian pinched vertex-correction diagrams for the $q\bar{q}\to C$ amplitude at one loop.  Particle notation is as defined in Fig. \protect\ref{fig:lsz}.}
\label{fig:NAVP}
\end{figure}

Of course we still need to include the contribution from $\Gamma_P^{\mu\nu\rho}(k,p,\bar{p})$ (of Eq. (\ref{eq:breakup2})) in the full non-Abelian vertex correction. This contains the pinching momenta: the action of $p$ and $\bar{p}$ on the external spinors gives zero, and the remaining piece cancels the internal fermion propagator in the diagram. Thus the internal fermion line in each diagram is ``pinched'' away, leaving an effective diagram with a 4-point coupling between fermions and gauge bosons as shown in Fig.~\ref{fig:NAVP}. The UV divergences of the pinched diagrams have the same group- and momentum-structure as those of the VPAs, and can be absorbed in the counterterms for the gauge field propagators. In order to see this clearly, we will now consider the form of the ``true'' propagator corrections to the $q\bar{q}\to C$ amplitude in the following subsection.

\subsection{Form of the Vacuum Polarization Amplitudes}
\label{sec:VPA}
\begin{figure}
\includegraphics[width=5.5in]{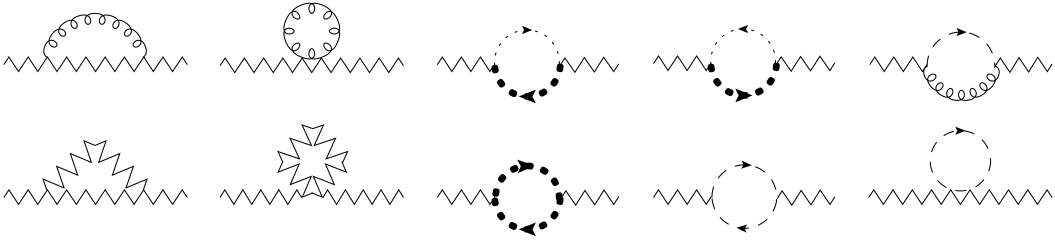}
\caption{Coloron-coloron vacuum polarization amplitude at one loop. A gluon field is, as usual, represented by a coiling line; a coloron field is represented by a zigzag line. The coloron ghost is represented by a sequence of filled circles, and the eaten Goldstone bosons are represented by dashed lines.}
\label{fig:CC}
\end{figure}
\begin{figure}
\includegraphics[width=5.5in]{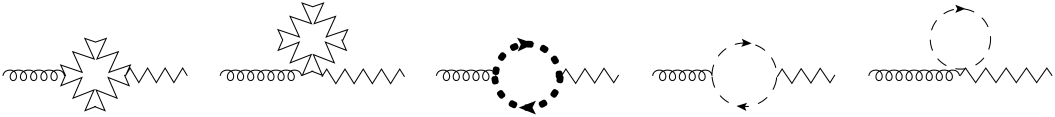}
\caption{Gluon-coloron mixing amplitude at one loop. Particle notation is as defined in Fig. \protect\ref{fig:CC}.}
\label{fig:GC}
\end{figure}
\begin{figure}[t]
\includegraphics[width=3.0in]{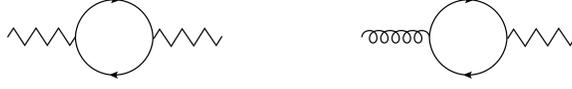}
\caption{Fermion contributions to coloron-coloron vacuum polarization amplitude and gluon-coloron mixing amplitude. Particle notation is as defined in Fig. \protect\ref{fig:CC}.}
\label{fig:CCGCfermions}
\end{figure}

The NLO corrections to the $q\bar{q}\to C$ amplitude due to the VPAs are found, from Eqs.~(\ref{eq:M}) and (\ref{eq:Z}), to have the form
\begin{eqnarray}
iP= i\mathcal{M}_{q\bar{q}\to C}^{(0)} \ \frac{\delta Z_C}{2} + i\mathcal{M}_{q\bar{q}\to C}^{\prime\prime(0)}
\frac{\Pi_{GC}(\hat{s})}{\hat{s}} \ ,
\label{eq:Pform}
\end{eqnarray}
where
\begin{eqnarray}
\delta Z_C = \Pi^\prime_{CC}(M^2_C) \ ,
\label{eq:derivative}
\end{eqnarray}
and
\begin{eqnarray}
i\mathcal{M}_{q\bar{q}\to C}^{\prime\prime(0)} = g_s\ \bar{v}^r(\bar{p}) \ i \gamma^{\mu}  t^a \ u^s (p) \ \varepsilon_\mu^{a\lambda\ast}(r) \ .
\end{eqnarray}
In order to obtain the second term of Eq.~(\ref{eq:Pform}), we have replaced $\Gamma^{a\mu}_{qqG}$ with its LO component $i\gamma^\mu t^a$. Notice also that at this order we can swap $M^2_{C {\rm phys}}$ for $M_C^2$.

At one loop $\Pi_{CC}(q^2)$ is given by the diagrams of Figs.~\ref{fig:CC} and \ref{fig:CCGCfermions}, in which the gluon ghost is represented by dotted lines, the coloron ghost by a sequence of filled circles, and the eaten Goldstone bosons are represented by dashed lines. There are poles at $d=2$ proportional to both $q^2$ and $M_C^2$. The latter correspond to quadratic divergences (renormalizing the coloron mass scale $f$), whereas the former can only be logarithmic by dimensional analysis (renormalizing the coloron field).\footnote{This situation parallels the renormalization of the electroweak chiral lagrangian \protect\cite{Appelquist:1980vg,Appelquist:1980ae}.} The momentum dependent part of the full coloron-coloron VPA is not transverse, as the coefficients of the $q^2$ and $q^\mu q^\nu$ terms are different. However we have explicitly verified that the infinite part is transverse: this is necessary, because the corresponding Lagrangian counterterms {\em are} transverse. For small values of $\epsilon$ we obtain
\begin{eqnarray}
&&\left(\frac{\alpha_s}{4\pi}\right)^{-1}\Pi_{CC}(q^2)g^{\mu\nu}+q^\mu q^\nu\ {\rm terms} = C_2(G)\int_0^1 dx \Bigg\{
\Bigg[\left(\frac{\mu^2}{\Delta_{GC}}\right)^\epsilon 2\Big(1+4x(1-x)\Big)E+2(1-2x)^2\Bigg]\left(g^{\mu\nu}q^2-q^\mu q^\nu\right) \nonumber \\
&&+\Bigg[\left(\frac{\mu^2}{\Delta_{GC}}\right)^\epsilon \Big(1-x(4-3x)\Big) E-x(1-x)\Bigg] g^{\mu\nu}q^2
+\Bigg[\left(\frac{\mu^2}{\Delta_{GC}}\right)^\epsilon 2 x\ E +3-5x\Bigg]g^{\mu\nu} M_C^2 \Bigg\} \nonumber \\
&&+4\cot^2(2\theta_c) C_2(G) \int_0^1 dx \Bigg\{
\Bigg[\left(\frac{\mu^2}{\Delta_{CC}}\right)^\epsilon \Big(1+4x(1-x)\Big)E+(1-2x)^2\Bigg]\left(g^{\mu\nu}q^2-q^\mu q^\nu\right) \nonumber \\
&&+\Bigg[-\left(\frac{\mu^2}{\Delta_{CC}}\right)^\epsilon \frac{x(1-x)}{4} E-\frac{x(1-x)}{4}\Bigg] g^{\mu\nu} q^2
+\left(\frac{\mu^2}{\Delta_{CC}}\right)^\epsilon \frac{(1-2x)^2}{8}\ E\ q^\mu q^\nu
+\Bigg[\left(\frac{\mu^2}{\Delta_{CC}}\right)^\epsilon \frac{5}{4}E+\frac{1}{4}\Bigg]g^{\mu\nu}M_C^2\Bigg\} \nonumber \\
&&+(r_L^2+r_R^2)N_f \int_0^1 dx \left(\frac{\mu^2}{\Delta_{qq}}\right)^\epsilon \Big[-2x(1-x)\Big] E\left(g^{\mu\nu}q^2-q^\mu q^\nu\right)~,
\label{eq:PCC}
\end{eqnarray}
where our results depend only on the coefficient of $g^{\mu\nu}$,  the quantity $N_f$ is the number of quark flavors in the loop (see
Fig. \ref{fig:CCGCfermions}),
\begin{equation}
E\equiv \frac{1}{\epsilon}-\gamma+\log 4\pi \ ,
\label{eq:E}
\end{equation}
and $\gamma$ is the Euler-Mascheroni constant. The $\Delta$ functions in Eq.~(\ref{eq:PCC}) are
\begin{eqnarray}
\Delta_{GC} &\equiv& x\ m_g^2+(1-x)M_C^2-x(1-x)q^2-i\eta~, \nonumber \\
\Delta_{CC} &\equiv& M_C^2-x(1-x)q^2-i\eta~, \nonumber \\
\Delta_{qq} &\equiv& -x(1-x)q^2 -i\eta \ .
\end{eqnarray}
Notice that the coloron-coloron VPA of Eq.~(\ref{eq:PCC}) is not IR divergent in the limit $m_g\to 0$, since there are no contributions with only massless (gluon) states. However what enters in Eq.~(\ref{eq:Pform}) is the {\em derivative} of $\Pi_{CC}$ (see Eq. (\ref{eq:derivative})), which {\em is} IR divergent in the limit $m_g\to 0$.

The momentum dependent part of the gluon-coloron mixing amplitude (Figs. \ref{fig:GC} and \ref{fig:CCGCfermions}) is found to be transverse, both in the infinite and the finite parts. For small values of $\epsilon$ we find
\begin{eqnarray}
\left(\frac{\alpha_s}{4\pi}\right)^{-1}\Pi_{GC}(q^2)g^{\mu\nu} &+& q^\mu q^\nu\ {\rm terms} \\
&=& 2 \cot(2\theta_c) C_2(G) \int_0^1 dx
\left(\frac{\mu^2}{\Delta_{CC}}\right)^\epsilon \Bigg\{
\Bigg[\Big(\frac{3}{4}+5x(1-x)\Big)E+(1-2x)^2\Bigg]
\left(q^2 g^{\mu\nu}-q^\mu q^\nu\right) +E\ M_C^2  \Bigg\} \nonumber \\
&+&(r_L+r_R)N_f \int_0^1 dx \left(\frac{\mu^2}{\Delta_{qq}}\right)^\epsilon
\Big[-2 x(1-x)\Big] \left(q^2 g^{\mu\nu}-q^\mu q^\nu\right)~. \nonumber
\end{eqnarray}
There are no potential IR divergences hidden in $\Pi_{GC}$.
\subsection{Non-Abelian Vertex Corrections {\em a la} Pinch-Technique: Pinched Diagrams} \label{sec:pinched}
The pinched diagrams of Fig.~\ref{fig:NAVP} are obtained from the diagrams of Fig.~\ref{fig:NAV} by replacing the full non-Abelian vertex momentum structure $\Gamma^{\mu\nu\rho}(k,p,\bar{p})$  from Eq. (\ref{eq:refa}), with $\Gamma_P^{\mu\nu\rho}(k,p,\bar{p})$ from Eq. (\ref{eq:breakup2}). This  leads to the amplitude
\begin{eqnarray}
i P_{\rm pinched} &=& \frac{\alpha_s}{4\pi} C_2(G) \int_0^1 dx \Bigg[
2\left(\frac{\mu^2}{\Delta_{GC}}\right)^\epsilon
+4\cot^2(2\theta_c) \left(\frac{\mu^2}{\Delta_{CC}}\right)^\epsilon
\Bigg] E\ \mathcal{M}_{q\bar{q}\to C}^{(0)} \nonumber \\
&+& \frac{\alpha_s}{4\pi}\ 2\cot(2\theta_c) C_2(G)\int_0^1 dx \left(\frac{\mu^2}{\Delta_{CC}}\right)^\epsilon E\
\mathcal{M}_{q\bar{q}\to C}^{\prime\prime(0)}~,
\label{eq:P}
\end{eqnarray}
where we have used Eq.~(\ref{eq:couplingrel}) to rewrite the fermion couplings in terms of $\theta_c$. This contribution to the amplitude has the form of a VPA correction, like that in Eq.~(\ref{eq:Pform}). In fact we can write
\begin{eqnarray}
iP_{\rm pinched}= i\mathcal{M}_{q\bar{q}\to C}^{(0)}  \frac{\widetilde{\Pi}^\prime_{CC}(M^2_C)}{2}
+ i\mathcal{M}_{q\bar{q}\to C}^{\prime\prime(0)} \frac{\widetilde{\Pi}_{GC}(\hat{s})}{\hat{s}} \ ,
\label{eq:Pformtilde}
\end{eqnarray}
where
\begin{eqnarray}
\left(\frac{\alpha_s}{4\pi}\right)^{-1}\widetilde{\Pi}_{CC}(q^2) &=& C_2(G) \int_0^1 dx
\left(\frac{\mu^2}{\Delta_{GC}}\right)^\epsilon 4(q^2-M_C^2)E \nonumber \\
&+&4\cot^2(2\theta_c) C_2(G) \int_0^1 dx \left(\frac{\mu^2}{\Delta_{CC}}\right)^\epsilon 2(q^2-M_C^2)E \ ,
\end{eqnarray}
and
\begin{eqnarray}
\left(\frac{\alpha_s}{4\pi}\right)^{-1}\widetilde{\Pi}_{GC}(q^2) = 2\cot(2\theta_c) C_2(G)\int_0^1 dx
\left(\frac{\mu^2}{\Delta_{CC}}\right)^\epsilon E\ q^2 ~.
\end{eqnarray}
\subsection{Full Propagator Correction}

We have just seen that, due to the pinch technique, the coloron-coloron and gluon-coloron VPAs receive an additional contribution from the pinched non-Abelian vertex corrections. Combining the VPAs, the UV divergences can be removed by two wavefunction renormalization counterterms (which arise from renormalizing the gauge eigenstates $G_{1\mu}$ and $G_{2\mu}$) and one mass counterterm (which arises from renormalizing the vacuum expectation value $f$), in the usual way. In the $\overline{\rm MS}$ scheme we obtain
\begin{eqnarray}
&&\left(\frac{\alpha_s}{4\pi}\right)^{-1}\Big[\Pi_{CC}(q^2)+\widetilde{\Pi}_{CC}(q^2)\Big] = C_2(G)\int_0^1 dx \Bigg\{
\Bigg[\left(\left(\frac{\mu^2}{\Delta_{GC}}\right)^\epsilon-1\right) 2\Big(3+4x(1-x)\Big)E+2(1-2x)^2\Bigg]q^2 \nonumber \\
&&+\Bigg[
\left(\frac{\mu^2}{\Delta_{GC}}\right)^\epsilon \Big(1-x(4-3x)\Big)E-x(1-x)\Bigg] q^2
+\Bigg[-\left(\left(\frac{\mu^2}{\Delta_{GC}}\right)^\epsilon-1\right)2(2-x)E +3-5x\Bigg] M_C^2 \Bigg\} \nonumber \\
&&+4\cot^2(2\theta_c) C_2(G) \int_0^1 dx \Bigg\{
\Bigg[\left(\left(\frac{\mu^2}{\Delta_{CC}}\right)^\epsilon-1\right) \Big(3+4x(1-x)\Big)E+(1-2x)^2\Bigg]q^2 \nonumber \\
&&-\Bigg[\left(\left(\frac{\mu^2}{\Delta_{CC}}\right)^\epsilon-1\right)E+1\Bigg]\frac{x(1-x)}{4}  q^2
+\Bigg[-\left(\left(\frac{\mu^2}{\Delta_{CC}}\right)^\epsilon-1\right) \frac{3}{4}E+\frac{1}{4}\Bigg]M_C^2\Bigg\} \nonumber \\
&&+(r_L^2+r_R^2)N_f \int_0^1 dx \left(\left(\frac{\mu^2}{\Delta_{qq}}\right)^\epsilon-1\right) \Big[-2x(1-x)\Big]E\ q^2 \ ,
\end{eqnarray}
and
\begin{eqnarray}
\left(\frac{\alpha_s}{4\pi}\right)^{-1}\Big[\Pi_{GC}(q^2) &+& \widetilde{\Pi}_{GC}(q^2)\Big] \\
&=&2 \cot(2\theta_c) C_2(G) \int_0^1 dx \Bigg\{
\left(\left(\frac{\mu^2}{\Delta_{CC}}\right)^\epsilon-1\right)
\Bigg[\Big(\frac{7}{4}+5x(1-x)\Big)q^2+M_C^2\Bigg]E +(1-2x)^2 q^2 \Bigg\} \nonumber \\
&+&(r_L+r_R)N_f \int_0^1 dx \
\left(\left(\frac{\mu^2}{\Delta_{qq}}\right)^\epsilon-1\right)\ \Big[-2x(1-x)\Big]E\ q^2 \ . \nonumber
\end{eqnarray}
The overall UV-finite propagator correction to the $q\bar{q}\to C$ amplitude can be found by insering these expressions in
\begin{eqnarray}
iP+iP_{\rm pinched} = i\mathcal{M}_{q\bar{q}\to C}^{(0)} \ \frac{\Pi^\prime_{CC}(M_C^2)+\widetilde{\Pi}^\prime_{CC}(M^2_C)}{2}
+ i\mathcal{M}_{q\bar{q}\to C}^{\prime\prime(0)} \displaystyle{\frac{\Pi_{GC}(\hat{s})+\widetilde{\Pi}_{GC}(\hat{s})}{\hat{s}}} \ .
\end{eqnarray}
Letting $m_g\to 0$, we find that $P+P_{\rm pinched}$ becomes IR divergent, with the divergence arising from $\Pi_{CC}^\prime$. Setting $m_g=0$ and $\epsilon<0$ gives
\begin{eqnarray}
iP+iP_{\rm pinched} = \frac{\alpha_s}{4\pi} C_2(G) \left(-\frac{1}{\epsilon}\right) i\mathcal{M}_{q\bar{q}\to C}^{(0)} + {\rm finite} \ .
\end{eqnarray}

\begin{figure}
\begin{center}
\includegraphics[width=4.5in]{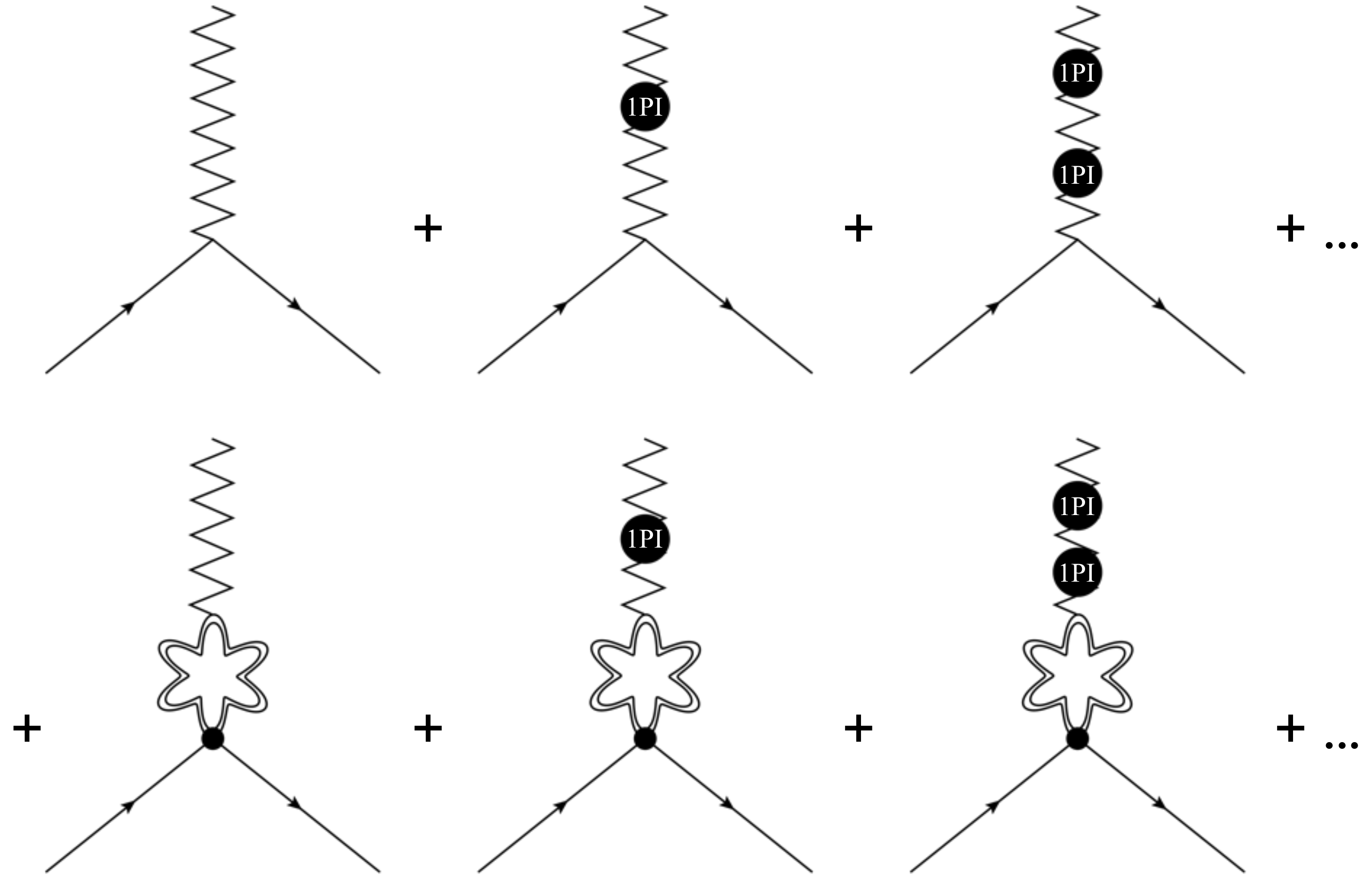}
\caption{The relevant contributions to the coloron Dyson series; as before, the zigzag lines represent colorons. The first row is the sum of the coloron VPA diagrams in the propagator, while the second row represents the sum of the VPA diagrams on top of the one-loop contribution from the ``pinched'' vertex correction (the double curly line illustrates generically all the allowed gauge bosons in the original non-Abelian vertices). The overall pinched amplitude factors out, and has no effect on the coloron pole mass.}
\label{fig:dyson}
\end{center}
\end{figure}
We have seen that the pinched diagrams contribute to the full propagators of the gluon-coloron system. This might seem in conflict with the expectation that the mass poles should be a property of freely propagating particles, and should not depend on any initial and/or final state. However, when we sum the Dyson series to obtain the full propagator, the pinched diagrams always appear as an overall pre-factor, as pictorially shown in Fig.~\ref{fig:dyson}. This has an overall effect on the full propagators, which depend on the initial and final states, but has no effect on the propagator poles. Thus when we compute physical masses, we can do so by employing the ``true'' propagators in the computation, without the contribution from the pinched diagrams.
\subsection{Cross Section at One Loop}

Adding up the tree-level contribution and the NLO contributions from $i Q+i V_{\rm Abelian}+i V_{\rm non-Abelian}$, and $iP+iP_{\rm pinched}$, gives a $q\bar{q}\to C$ amplitude of the form
\begin{eqnarray}
i\mathcal{M}_{q\bar{q}\to C} &=& i \mathcal{M}_{q\bar{q}\to C}^{(0)} + i Q+i V_{\rm Abelian}
+i V_{\rm non-Abelian} + iP+iP_{\rm pinched} \nonumber \\
&\equiv& i \mathcal{M}_{q\bar{q}\to C}^{(0)} + \frac{\alpha_s}{4\pi}\Big(
T\ i \mathcal{M}_{q\bar{q}\to C}^{(0)} + T^\prime\ i \mathcal{M}_{q\bar{q}\to C}^{\prime(0)}
+ T^{\prime\prime}\ i \mathcal{M}_{q\bar{q}\to C}^{\prime\prime(0)}
\Big) \ ,
\end{eqnarray}
where expressions for the real parts of $T$, $T'$, and $T''$ are given below.
Averaging the squared amplitude over initial spins and colors, summing over final polarization states, and integrating over the phase space, gives the NLO result of the form
\begin{eqnarray}
\hat{\sigma}_{\rm virt} \equiv\hat{\sigma}_{q\bar{q}\to C}^{(0)}+\hat{\sigma}_{q\bar{q}\to C}^{(1)} = \frac{\alpha_s A (r_L^2+r_R^2)}{\hat{s}}\ \delta(1-\chi) \Bigg[
1+\frac{\alpha_s}{2\pi}\Bigg(
{\rm Re}\ T + \frac{r_L^4+r_R^4}{r_L^2+r_R^2}\ {\rm Re}\ T^\prime + \frac{r_L+r_R}{r_L^2+r_R^2}\ {\rm Re}\ T^{\prime\prime}\Bigg)
\Bigg] \ .
\label{eq:virtual}
\end{eqnarray}
At $\hat{s}=M_C^2$ it is possible to integrate over the Feynman parameter space in the expressions for $i Q+i V_{\rm Abelian}+i V_{\rm non-Abelian}$, and $iP+iP_{\rm pinched}$. As we have seen, the UV infinities cancel in $i Q+i V_{\rm Abelian}+i V_{\rm non-Abelian}$ and are absorbed by propagator conterterms in $iP+iP_{\rm pinched}$. Thus for $m_g\neq 0$ the overall amplitude is finite. Taking the $m_g\to 0$ limit leads to IR divergences in ${\rm Re}\ T$, which are parametrized by taking $\epsilon<0$. For small and negative values of $\epsilon$ we obtain
\begin{eqnarray}
{\rm Re}\ T &=& \left(\frac{4\pi\mu^2}{M_C^2}\right)^{\epsilon} \Gamma(1+\epsilon)
\Bigg[
-\frac{2}{\epsilon^2}-\frac{3}{\epsilon}-8+\frac{4\pi^2}{3}
\Bigg]C_2(r)
+\Bigg[-E+\frac{61}{9}-\frac{5\pi}{2\sqrt{3}}-\frac{\pi^2}{3}-\frac{8}{3}\log\frac{M_C^2}{\mu^2}\Bigg]C_2(G)
\nonumber \\
&+&\Bigg[
\frac{77}{48}-\frac{7\pi}{16\sqrt{3}}-\frac{29}{16}\log\frac{M_C^2}{\mu^2}\Bigg]4\cot^2(2\theta_c) C_2(G)
+ \Bigg[-\frac{1}{9}+\frac{1}{6}\log\frac{M_C^2}{\mu^2}\Bigg](r_L^2+r_R^2)N_f  \ , \nonumber \\
{\rm Re}\ T^\prime &=&\Bigg[
-\frac{11}{2}+\frac{2\pi^2}{3}
\Bigg]C_2(r)
+\Bigg[
1+\frac{5\pi}{2\sqrt{3}}-\frac{2\pi^2}{3}
\Bigg]C_2(G) \ , \nonumber \\
{\rm Re}\ T^{\prime\prime} &=& \Bigg[\frac{95}{9}-\frac{7\sqrt{3}\pi}{4}-\frac{43}{12}\log\frac{M_C^2}{\mu^2}\Bigg]2\cot\left(2\theta_c\right)C_2(G)
+\Bigg[-\frac{5}{9}+\frac{1}{3}\log\frac{M_C^2}{\mu^2}\Bigg](r_L+r_R)N_f~.
\label{eq:T}
\end{eqnarray}
In the next section we will compute the corrections to the tree-level cross section due to the emission of soft and collinear gluons. We will show that the real emission cross section has IR divergences which exactly cancel the IR divergences contained in $\hat{\sigma}_{\rm virt}$ (Eq. (\ref{eq:virtual})), leading to a total cross-section free of both UV and IR divergences.
\section{NLO Coloron Production: Real Corrections} \label{sec:real}
\begin{figure}
\begin{center}
\includegraphics[width=5in]{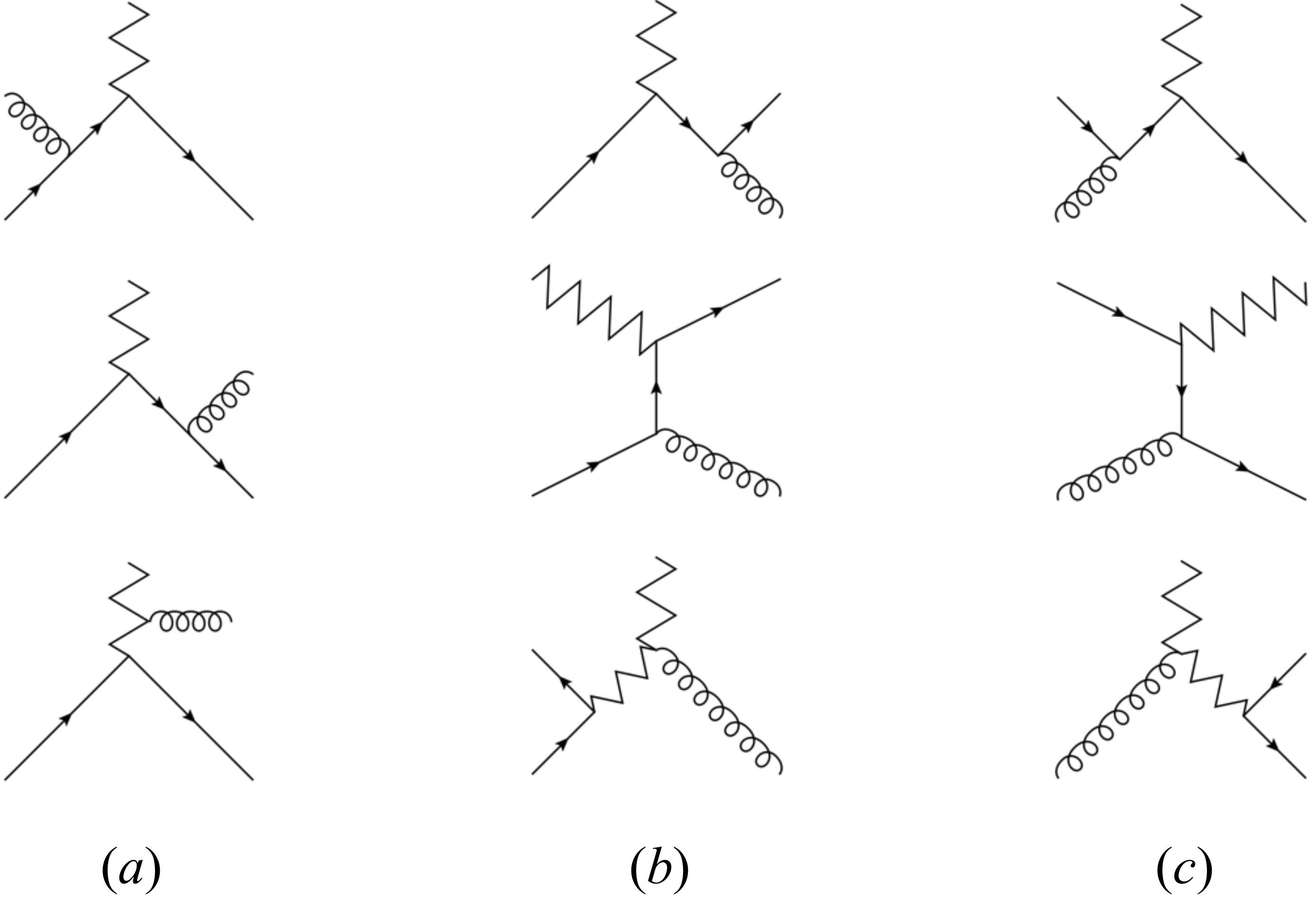}
\caption{Diagrams contributing to the real emission processes. A gluon field is, as usual, represented by a coiling line; a coloron field is represented by a zigzag line. $(a)$ Gluon emission. $(b)$ Quark emission. $(c)$ Antiquark emission.}
\label{fig:real}
\end{center}
\end{figure}
The real emission corrections, at NLO, are given by the diagrams of Fig~\ref{fig:real}. We first consider the diagrams with real emission of a gluon, shown in Fig~\ref{fig:real} (a). The squared amplitude, averaged over initial colors and spins, and summed over final colors and polarizations, is found to be, in $d=2(2-\epsilon)$ dimensions,
\begin{align}
&\overline{| \mathcal{M}_{q\bar{q} \rightarrow gC}^{(1)} |^{2}} = \, \frac{C_{2}(r)\, g_{s}^{4} \, (r_{L}^{2}+r_{R}^{2}) }{{\rm dim}(r)}\, \mu^{2\epsilon} \, (1-\epsilon) \left[ \frac{-1}{\omega(1-\omega)} \, C_{2}(r) + C_{2}(G) \right] \left[ \epsilon - \frac{1+\chi^{2}}{(1-\chi)^{2}} + 2\omega \, (1-\omega) \right]  \ . \label{MRgsqavfin} 
\end{align}
where
\begin{equation}
\omega\equiv \frac{1-\cos\theta}{2} \ ,
\label{eq:omega}
\end{equation}
$\chi$ was defined in Eq. (\ref{eq:chi}), and $\theta$ is the angle between the emitted gluon and the colliding quarks. The cross section for the real gluon emission is
\begin{equation} \label{csRg}
\hat{\sigma}_{q\bar{q}\to g C}^{(1)} = \frac{1}{2\hat{s}} \int d\Pi_{2} \, \overline{|\mathcal{M}_{q\bar{q}\to g C}^{(1)}|^{2}} \ ,
\end{equation}
where the integral is over the 2-body Lorentz-invariant phase space in parton CM. In $d=4-2\epsilon$ dimensions,
\begin{equation} \label{LIPSd}
\int d\Pi_{2} = \frac{1}{8\pi} \, \frac{1-\chi}{\Gamma(1-\epsilon)} \left[ \frac{M_C^2(1-\chi)^{2}}{4\pi \chi} \right]^{-\epsilon} \int_{0}^{1}d \omega \left[ \omega \, (1-\omega) \right]^{-\epsilon} \ .
\end{equation}
This leads to the partonic cross section
\begin{eqnarray}
\hat{\sigma}_{q\bar{q}\to g C}^{(1)} = \frac{\alpha_s (r_L^2+r_R^2) A}{\hat{s}}\ \frac{\alpha_s}{2\pi}
\left(\frac{4\pi \mu^{2}}{M_C^2} \right)^\epsilon
\frac{\Gamma(1-\epsilon)}{\Gamma(1-2\epsilon)}
\Bigg[-C_2(r) \, \frac{2}{\epsilon} \frac{\chi^\epsilon(1 + \chi^2)}{(1-\chi)^{1+2\epsilon}}
- C_2(G) \  \frac{2}{3}\frac{\chi^\epsilon(1+\chi+\chi^2)}{(1-\chi)^{1+2\epsilon}} \Bigg] \ .
\label{eq:csreal}
\end{eqnarray}
Now $\chi$ is no longer constrained to be equal to one. Instead we must have $\chi\leq 1$, or else no on-shell coloron can be produced. The term proportional to $C_2(r)$ features a collinear singularity, parametrized by $\epsilon$, and a soft singularity, parametrized by $1-\chi$. The term proportional to $C_2(G)$ only features a soft singularity. The integral over $\chi$ in Eq.~(\ref{eq:csreal}) is finite for $\epsilon<0$, in spite of the singularity of the integrands.
For small and negative values of $\epsilon$ we can rewrite the $\chi$-dependence as follows
\begin{eqnarray}
\frac{\chi^\epsilon(1 + \chi^2)}{(1-\chi)^{1+2\epsilon}} &=& -\frac{1}{\epsilon}\delta(1-\chi)
+\frac{1+\chi^2}{(1-\chi)_+}-\Bigg[2(1+\chi^2)\left(\frac{\log(1-\chi)}{1-\chi}\right)_{+}
-\frac{1+\chi^2}{1-\chi}\log\chi\Bigg] \epsilon~, \nonumber \\
\frac{\chi^\epsilon(1+\chi+\chi^2)}{(1-\chi)^{1+2\epsilon}}  &=& - \frac{3}{2\epsilon}\delta(1-\chi)
+\frac{1+\chi+\chi^2}{(1-\chi)_+}~,
\end{eqnarray}
where, as conventional, the ``+'' distributions are defined by
\begin{eqnarray}
\int_0^1 d\chi \frac{f(\chi)}{(1-\chi)_+} \equiv \int_0^1 d\chi \frac{f(\chi)-f(1)}{1-\chi} \ , \quad
\int_0^1 d\chi\ f(\chi)\left(\frac{\log(1-\chi)}{1-\chi}\right)_{+} \equiv
\int_0^1 d\chi \left[f(\chi)-f(1)\right]\frac{\log(1-\chi)}{1-\chi} \ .
\end{eqnarray}
The coefficients of the delta functions are found by integrating both sides of the equations. The partonic cross section becomes
\begin{eqnarray}
\hat{\sigma}_{q\bar{q}\to g C}^{(1)} =
\frac{\alpha_s A (r_L^2+r_R^2)}{\hat{s}}\ \frac{\alpha_s}{2\pi}
\Bigg[\delta(1-\chi)\ R + R^\prime\Bigg] \ ,
\label{eq:csrealsplit}
\end{eqnarray}
where, using Eq.~(\ref{eq:E}), and expanding for small values of $\epsilon$,
\begin{eqnarray}
R &=& \left(\frac{4\pi\mu^2}{M_C^2}\right)^\epsilon \frac{\Gamma(1-\epsilon)}{\Gamma(1-2\epsilon)}\Bigg\{
C_2(r)\Bigg[\frac{2}{\epsilon^2}+\frac{3}{\epsilon}\Bigg] + C_2(G)\frac{1}{\epsilon}\Bigg\}\ , \nonumber \\
R^\prime &=&-2\Bigg[E-\log\frac{M_C^2}{\mu^2}\Bigg] P_{q\to q}(\chi)
+C_2(r)\Bigg[4(1+\chi^2)\left(\frac{\log(1-\chi)}{1-\chi}\right)_+ -2\frac{1+\chi}{1-\chi}\log\chi\Bigg]
+C_2(G)\frac{2}{3}\frac{1+\chi+\chi^2}{(1-\chi)_+}  \ .
\label{eq:RRp}
\end{eqnarray}
In the second equation $P_{q\to q}(\chi)$ is the Altarelli-Parisi splitting function for an on-shell quark to evolve into a virtual quark and a real gluon:
\begin{eqnarray}
P_{q\to q}(\chi) = C_2(r)\Bigg[\frac{1+\chi^2}{(1-\chi)_+}+\frac{3}{2}\delta(1-\chi)\Bigg] \ .
\end{eqnarray}
Adding together $\hat{\sigma}_{\rm virt}$, given by Eqs.~(\ref{eq:virtual}) and (\ref{eq:T}), and $\hat{\sigma}_{q\bar{q}\to gC}^{(1)}$, given by Eq. (\ref{eq:csrealsplit}), shows that the IR divergences proportional to $\delta(1-\chi)$ cancel. There is still a collinear singularity in $R^\prime$, proportional to the Altarelli-Parisi evolution $P_{q\to q}(\chi)$. This singularity arises from integrating over all collinear initial-state gluons.
As we will see in the next section, these collinear IR divergences will be absorbed through renormalization of the PDFs.

The real quark and antiquark emission diagrams are shown in Fig.~\ref{fig:real} (b) and (c), respectively. The corresponding summed-averaged squared amplitudes are
\begin{align}
&\overline{| \mathcal{M}_{qg \rightarrow qC}^{(1)} |^{2}} = \frac{C_{2}(r)\, g_{s}^{4} \, (r_{L}^{2}+r_{R}^{2}) }{{\rm dim}(G)}\, \mu^{2\epsilon} \label{MRfsqavfin} \\
& \times \left[ C_{2}(r) + C_{2}(G) \frac{(1-\chi)(1-\omega)}{(1-(1-\chi)(1-\omega))^2} \right] \left[ 2\,(\epsilon+\chi) + \frac{1 - \epsilon -2\chi(1- \chi)}{(1 - \chi) \, (1-\omega)} +  (1-\epsilon)(1-\chi) \, (1-\omega) \right] \notag \ .
\end{align}
and
\begin{align}
&\overline{| \mathcal{M}_{\bar{q}g \rightarrow \bar{q}C}^{(1)} |^{2}} = \frac{C_{2}(r)\, g_{s}^{4} \, (r_{L}^{2}+r_{R}^{2}) }{{\rm dim}(G)}\, \mu^{2\epsilon} \label{MRfbsqavfin} \\
& \times \left[ C_{2}(r) + C_{2}(G) \frac{(1-\chi)\, \omega}{(1-(1-\chi)\, \omega)^2} \right] \left[ 2\,(\epsilon+\chi) + \frac{1 - \epsilon -2\chi(1- \chi)}{(1 - \chi) \, \, \omega} +  (1-\epsilon)(1-\chi) \, \, \omega \right] \notag \ ,
\end{align}
where ${\rm dim}(G)\equiv 8$ is the dimension of the adjoint representation. Note that the amplitudes for quark
and antiquark emission are related by crossing, {\it i.e.} $\omega \leftrightarrow (1-\omega)$.
The integration over the two-body Lorentz-invariant phase space proceeds as in the gluon emission case, yielding
\begin{eqnarray}
\hat{\sigma}_{q g\to q C}^{(1)} = \hat{\sigma}_{\bar{q} g\to \bar{q} C}^{(1)} &=& 
\frac{\alpha_s A (r_L^2+r_R^2)}{\hat{s}}\ \frac{\alpha_s}{2\pi}\ R^{\prime\prime} \ ,
\label{eq:refextra}
\end{eqnarray}
where
\begin{eqnarray}
R^{\prime\prime} &=& \frac{{\rm dim}(r)}{{\rm dim}(G)}
\Bigg\{C_2(r)\frac{3+2\chi-3\chi^2}{2}
+C_2(G)\Bigg[\frac{(1-\chi)(2+\chi+2\chi^2)}{\chi}+2(1+\chi)\log\chi\Bigg]\Bigg\} \nonumber \\
&-&\Bigg[E-\log\frac{M_C^2}{\mu^2}-\log\frac{(1-\chi)^2}{\chi}+1\Bigg]P_{g\to q}(\chi) \ .
\label{eq:Rpp}
\end{eqnarray}
Here $P_{g\to q}(\chi)$ is the Altarelli-Parisi splitting function for an on-shell gluon to evolve to a virtual-real quark pair,
\begin{eqnarray}
P_{g\to q}(\chi) =\frac{C_2(r)\cdot{\rm dim}(r)}{{\rm dim}(G)} \left[\chi^2+(1-\chi)^2\right] \ ,
\end{eqnarray}
where $C_2(r)\cdot {\rm dim}(r)/{\rm dim}(G)= 1/2$. There is no soft singularity in $\hat{\sigma}_{q g\to q C}^{(1)}\equiv \hat{\sigma}_{\bar{q} g\to \bar{q} C}^{(1)}$, only a collinear singularity proportional to the Altarelli-Parisi evolution $P_{g\to q}(\chi)$. As noted above regarding $\hat{\sigma}_{q\bar{q}\to g C}^{(1)}$, this singularity will be canceled by renormalization of the PDFs when we compute the
total hadronic cross section.
\section{NLO Cross Section}\label{sec:NLOcs}

Our calculations in the previous sections have produced all of the relevant partonic cross sections at NLO and demonstrated them to be both UV and IR finite.   Note that the  $g g\to C$ process vanishes at tree level \cite{Chivukula:2001gv} and the one-loop contributions are small, less than of order 0.1\% of the $q\bar{q}$-initiated
leading order contribution \cite{Allanach:2009vz}; we therefore do not include it in this work.

 The full NLO cross section for coloron production at the LHC is
\begin{eqnarray}
\sigma^{NLO} &=& \int d x_1 \int d x_2 \Bigg\{
\sum_q\Bigg[f^0_q(x_1) f^0_{\bar{q}}(x_2) + f^0_{\bar{q}}(x_1) f^0_q(x_2)\Bigg]\Big(\hat{\sigma}_{q\bar{q}\to C}^{(0)}
+\hat{\sigma}_{q\bar{q}\to C}^{(1)} + \hat{\sigma}_{q\bar{q}\to g C}^{(1)}\Big)  \nonumber \\
&+& \sum_q\Bigg[f^0_q(x_1) f^0_g(x_2)
+ f^0_g(x_1) f^0_q(x_2)+ f^0_{\bar{q}}(x_1) f^0_g(x_2) + f^0_g(x_1) f^0_{\bar{q}}(x_2)\Bigg] \hat{\sigma}_{q g\to q C}^{(1)}
\Bigg\} \ ,
\label{eq:barePDFcs}
\end{eqnarray}
where the partonic cross-sections $\hat{\sigma}$ are given in Eqs. (\ref{cstree}), (\ref{eq:virtual}), (\ref{eq:csrealsplit}), and (\ref{eq:refextra}), and
where the superscript ``0'' in the PDFs will be clear in a moment. We saw that all IR divergences contained in $\sigma$ cancel, except for a couple of collinear singularities proportional to Altarelli-Parisi evolutions. Such singularities arise because we integrated over all collinear quarks and gluons, even those which we should have included in the PDFs. Therefore, the corresponding IR singularities are absorbed by renormalizing the bare PDFs in Eq.~(\ref{eq:barePDFcs}). In the $\overline{{\rm MS}}$ scheme,
\begin{equation}
f_i(x, \mu_F) = f^0_i (x) - \frac{g_3^2}{8\pi^2} \brac{\frac{1}{\epsilon} - \gamma + \log (4\pi) - \log \frac{\mu_F^2}{\mu^2}} \int \frac{d\chi}{\chi} \sum_j \, f^0_j \brac{\frac{x}{\chi}} P_{j \rightarrow i} (\chi) \ ,
\label{eq:renormPDF}
\end{equation}
where $i,j=q,g$, and $\mu_F$ is the factorization scale. Exchanging the bare PDFs for the renormalized ones replaces $E$ with $\log\mu_F^2/\mu^2$ in Eqs.~(\ref{eq:RRp}) and (\ref{eq:Rpp}). The hadronic cross section becomes
\begin{eqnarray}
\sigma^{NLO} = \frac{\alpha_s\ A\ H_1(\theta_c)}{s}&\cdot&\int \frac{d x_1}{x_1} \int \frac{d x_2}{x_2} \Bigg\{
\sum_q\Bigg[f_q(x_1,\mu_F) f_{\bar{q}}(x_2,\mu_F)+f_{\bar{q}}(x_1,\mu_F)f_q(x_2,\mu_F)\Bigg]
\Bigg(\delta(1-\chi)+\frac{\alpha_s}{2\pi}\ {\cal F}^{qq}(\chi)\Bigg) \nonumber \\
&+& \sum_q\Bigg[f_q(x_1,\mu_F)  f_g(x_2, \mu_F) + f_g(x_1, \mu_F) f_q(x_2,\mu_F)
+( f_q \to f_{\bar{q}})
\Bigg] \frac{\alpha_s}{2\pi}\ {\cal F}^{qg}(\chi)\Bigg\} \ , 
\end{eqnarray}
where the function $H_1(\theta_c)$ is defined below, in Eq.~(\ref{eq:H}), $A$ is defined in Eq. (\ref{eq:defA}), and the partonic CM energy $\hat{s}$ has been traded for the hadronic one, as in Eq.~(\ref{eq:x1x2}). Notice that since the integrand is now finite, we can ignore the $1-\epsilon$ factor in $A$. The functions ${\cal F}^{qq}(\chi)$ and ${\cal F}^{qg}(\chi)$ are
\begin{eqnarray}
{\cal F}^{qq}(\chi) &=& 2\log\frac{M_C^2}{\mu_F^2} P_{q\to q}(\chi) +D_q(\chi) \ , \nonumber \\
{\cal F}^{qg}(\chi) &=& \log\frac{M_C^2}{\mu_F^2} P_{g\to q}(\chi) +D_g(\chi) \ ,
\label{eq:F}
\end{eqnarray}
where
\begin{eqnarray}
D_q(\chi) &=& C_2(r)\Bigg[4(1+\chi^2)\left(\frac{\log(1-\chi)}{1-\chi}\right)_+ -2\ \frac{1+\chi}{1-\chi}\log\chi\Bigg]
+C_2(G)\ \frac{2}{3}\frac{1+\chi+\chi^2}{(1-\chi)_+}+\delta(1-\chi)\ Q  \ ,\nonumber \\
D_g(\chi) &=& \frac{{\rm dim}(r)}{{\rm dim}(G)}\Bigg\{
C_2(r)\Bigg[\Big(\chi^2+(1-\chi)^2\Big)\Bigg(\log\frac{(1-\chi)^2}{\chi}-1\Bigg)+\frac{3}{2}+\chi-\frac{3}{2}\chi^2\Bigg]
+C_2(G)\Bigg[\frac{(1-\chi)(2+\chi+2\chi^2)}{\chi} \nonumber \\
&+&2(1+\chi)\log\chi\Bigg]
\Bigg\} \ ,
\end{eqnarray}
and
\begin{eqnarray}
Q &=& N_f\Bigg[\Bigg(-\frac{1}{9}+\frac{1}{6}\log\frac{M_C^2}{\mu^2}\Bigg)H_1(\theta_c)
+ \Bigg(-\frac{5}{9}+\frac{1}{3}\log\frac{M_C^2}{\mu^2}\Bigg)H_2(\theta_c)\Bigg]
+C_2(r)\Bigg[-8+\frac{2\pi^2}{3}+\Bigg(-\frac{11}{2}+\frac{2\pi^2}{3}\Bigg)H_3(\theta_c)\Bigg] \nonumber \\
&+&C_2(G)\Bigg[\frac{61}{9}-\frac{5\pi}{2\sqrt{3}}-\frac{\pi^2}{3}-\frac{11}{3}\log\frac{M_C^2}{\mu^2}
+\Bigg(\frac{77}{12}-\frac{7\pi}{4\sqrt{3}}-\frac{29}{4}\log\frac{M_C^2}{\mu^2}\Bigg)\cot^2(2\theta_c)
+\Bigg(1+\frac{5\pi}{2\sqrt{3}}-\frac{2\pi^2}{3}\Bigg)H_3(\theta_c) \nonumber \\
&+&\Bigg(\frac{190}{9}-\frac{7\sqrt{3}\pi}{2}-\frac{43}{6}\log\frac{M_C^2}{\mu^2}\Bigg)\cot(2\theta_c)H_4(\theta_c)
\Bigg] \ .
\label{eq:QNLO}
\end{eqnarray}
The functions $H_i(\theta_c)$ are determined by the chiral couplings of the quarks to the colorons (which depend on the charges of the quarks under the full $SU(3)_{1c} \times SU(3)_{2c}$ symmetry):
\begin{eqnarray}
&&H_1(\theta_c) = \left\{
\begin{array}{lr}
2\tan^2\theta_c & r_L=r_R=-\tan\theta_c \\
\ & \  \\
\tan^2\theta_c + \cot^2\theta_c & r_L\neq r_R \\
\ & \  \\
2\cot^2\theta_c & r_L=r_R=\cot\theta_c
\end{array}
\right. \ ,\ \
H_2(\theta_c) = \left\{
\begin{array}{lr}
2 & r_L=r_R=-\tan\theta_c \\
\ & \  \\
\displaystyle{\frac{2(1+\cos(4\theta_c))}{3+\cos(4\theta_c))}} & r_L\neq r_R \\
\ & \  \\
2 & r_L=r_R=\cot\theta_c
\end{array}
\right.  \ , \nonumber \\
&&H_3(\theta_c) = \left\{
\begin{array}{lr}
\tan^2\theta_c & r_L=r_R=-\tan\theta_c \\
\ & \  \\
\displaystyle{\frac{\tan^4\theta_c+\cot^4\theta_c}{\tan^2\theta_c+\cot^2\theta_c}} & r_L\neq r_R \\
\ & \  \\
\cot^2\theta_c & r_L=r_R=\cot\theta_c
\end{array}
\right. \ ,\ \
H_4(\theta_c) = \left\{
\begin{array}{lr}
-\cot\theta_c & r_L=r_R=-\tan\theta_c \\
\ & \  \\
\displaystyle{\frac{\sin(4\theta_c)}{3+\cos(4\theta_c)}} & r_L\neq r_R \\
\ & \  \\
\tan\theta_c & r_L=r_R=\cot\theta_c
\end{array}
\right. \ .
\label{eq:H}
\end{eqnarray}

At NLO the $\mu$ dependence is removed by trading the $\overline{\rm MS}$ couplings $g_{1s}$ and $g_{2s}$, or $g_s$ and $\theta_c$, for the corresponding running couplings. Since $\theta_c$ is a free parameter, we simply set $\mu\equiv M_C$, and express the cross section as a function of the $\overline{\rm MS}$ couplings. At the same time,
the NLO $\mu_F$ dependence weakens once the renormalized PDFs are employed, as $\sigma$ in Eq.~(\ref{eq:barePDFcs}) is independent of $\mu_F$
to this order in perturbation theory.

From these results we may also compute the transverse momentum distribution of the produced coloron, which  is given by
\begin{eqnarray}
\frac{d\sigma}{d p_T} &=& \int d x_1 \int d x_2 \Bigg\{
\sum_q \Bigg[f_q(x_1,\mu_F) f_{\bar{q}}(x_2,\mu_F) + f_{\bar{q}}(x_1,\mu_F) f_q(x_2,\mu_F)\Bigg]
\frac{d\hat{\sigma}_{q\bar{q}\to g C}}{d p_T}  \label{eq:dsigmadpt}\\
+\sum_q\Bigg[f_q(x_1,\mu_F) &f_g&(x_2,\mu_F)+ f_g(x_1,\mu_F) f_q(x_2,\mu_F)+ f_{\bar{q}}(x_1,\mu_F) f_g(x_2,\mu_F) + f_g(x_1,\mu_F) f_{\bar{q}}(x_2,\mu_F)\Bigg]
\frac{d\hat{\sigma}_{q g\to q C}}{d p_T}
\Bigg\} \ ,
\nonumber
\end{eqnarray}
where
\begin{eqnarray}
&&\frac{d\hat{\sigma}_{q\bar{q}\to g C}}{d p_T} = \frac{1}{4\pi\hat{s}^2 (1-\chi^2)}
\frac{p_T}{\displaystyle{\sqrt{1-\frac{4p_T^2}{\hat{s}(1-\chi)^2}}}}
\cdot 2 \overline{|\mathcal{M}_{q\bar{q}\to g C}^{(1)}|^{2}} \ , \\
&&\frac{d\hat{\sigma}_{q g\to q C}}{d p_T} = \frac{1}{4\pi\hat{s}^2 (1-\chi^2)}
\frac{p_T}{\displaystyle{\sqrt{1-\frac{4p_T^2}{\hat{s}(1-\chi)^2}}}}
\Bigg( \overline{|\mathcal{M}_{q g\to q C}^{(1)}|^{2}} + \overline{|\mathcal{M}_{q g\to q C}^{(1)}|^{2}}_{\omega\to 1-\omega} \Bigg) \ ,
\end{eqnarray}
and $\omega$ (Eq. (\ref{eq:omega})) is given by
\begin{eqnarray}
\omega=\frac{1-\displaystyle{\sqrt{1-\frac{4p_T^2}{\hat{s}(1-\chi)^2}}}}{2} \ .
\end{eqnarray}
Note that this is the {\it leading order} prediction for $d\sigma/dp_T$, and therefore this distribution is strongly
$\mu_F$-dependent.
\section{Discussion}\label{sec:conclusion}

\begin{figure}
\includegraphics[width=3.0in]{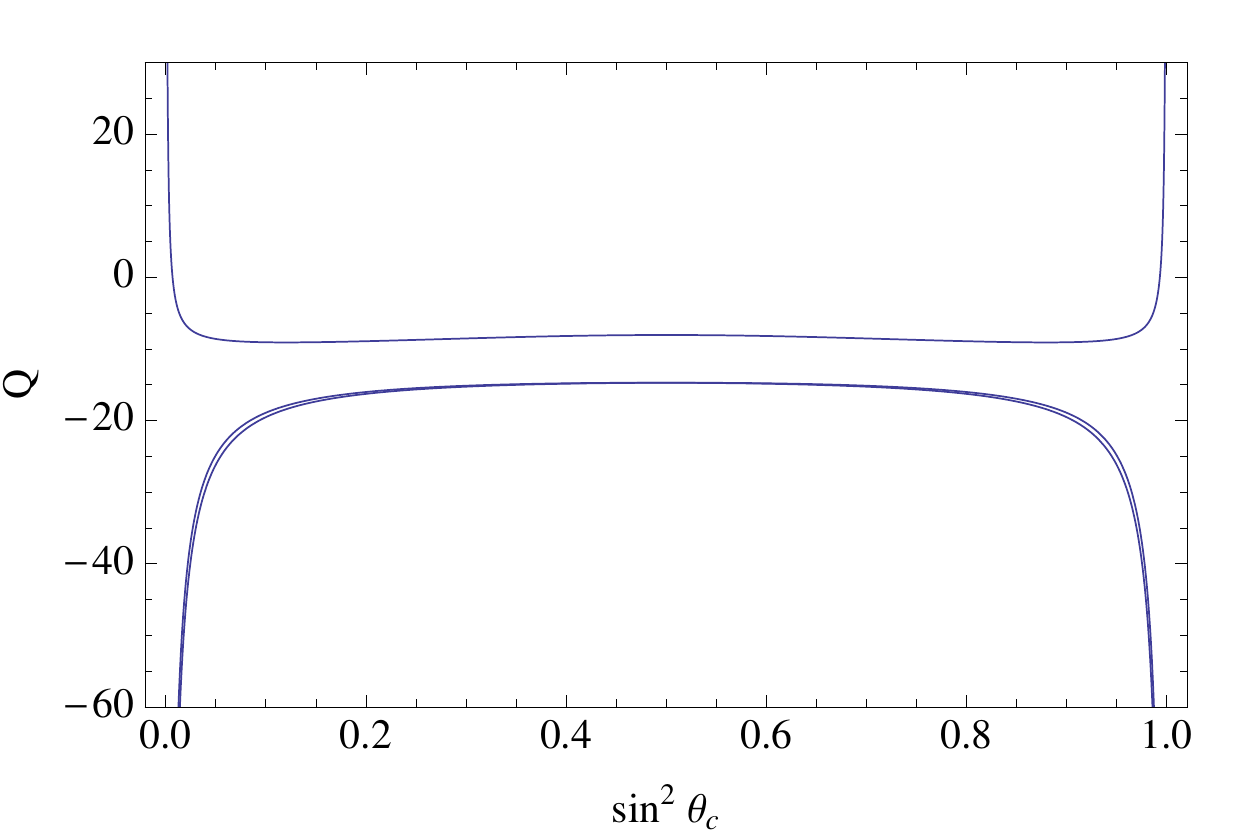}
\caption{Behavior of the $Q$ function defined in Eq.~(\ref{eq:QNLO}), for $\mu=M_C$: this gives the contribution from the virtual corrections to the NLO cross section for coloron production. The upper curve is for the $r_L\neq r_R$ scenario, whereas the almost identical lower curves are for $r_L=r_R=-\tan\theta_c$, and $r_L=r_R=\cot\theta_c$.
Note that $Q$, and therefore the NLO corrections, become very large when $\sin^2\theta_c$ is either too small or too large.}
\label{fig:Q}
\end{figure}

\begin{figure}
\includegraphics[width=2.3in]{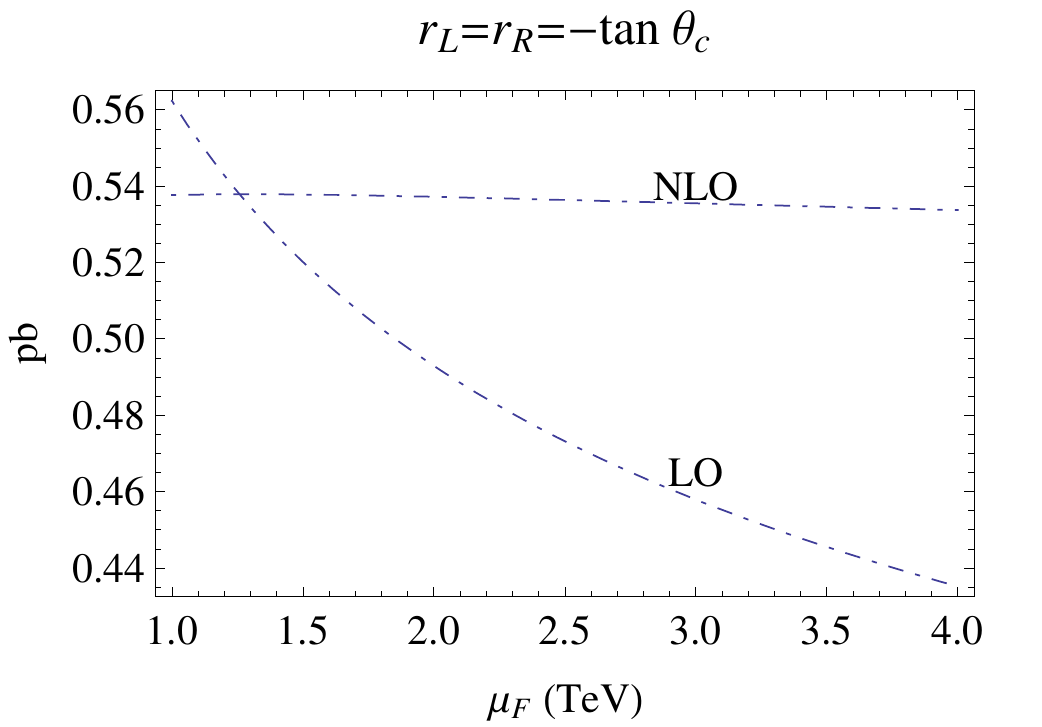}
\includegraphics[width=2.3in]{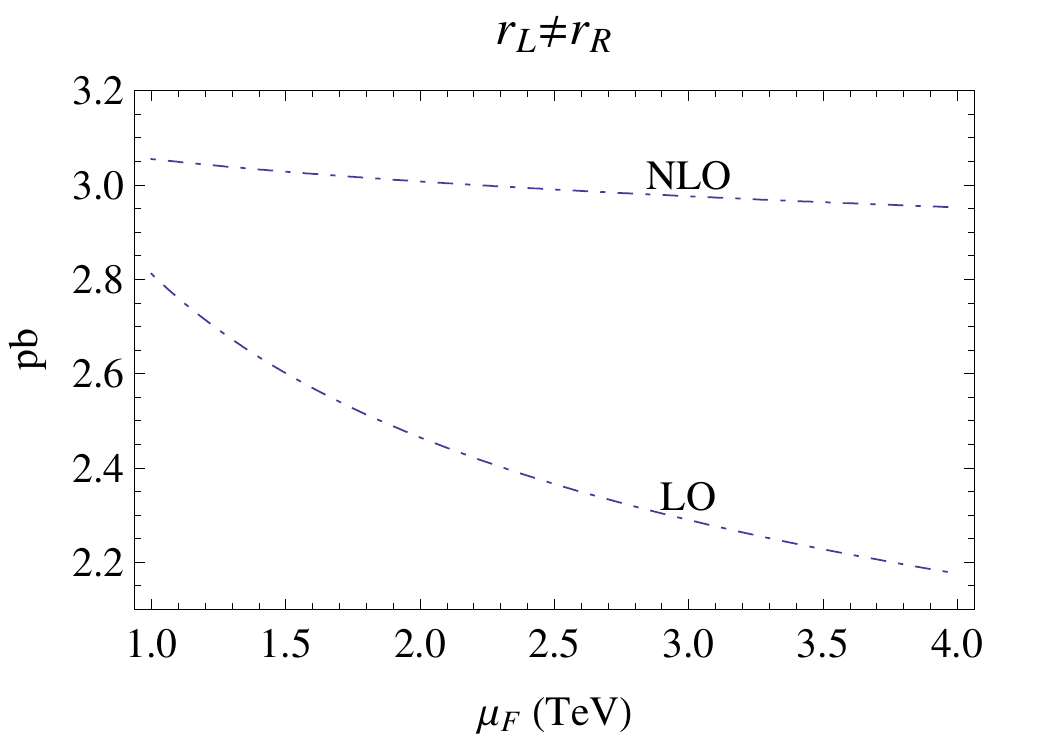}
\includegraphics[width=2.3in]{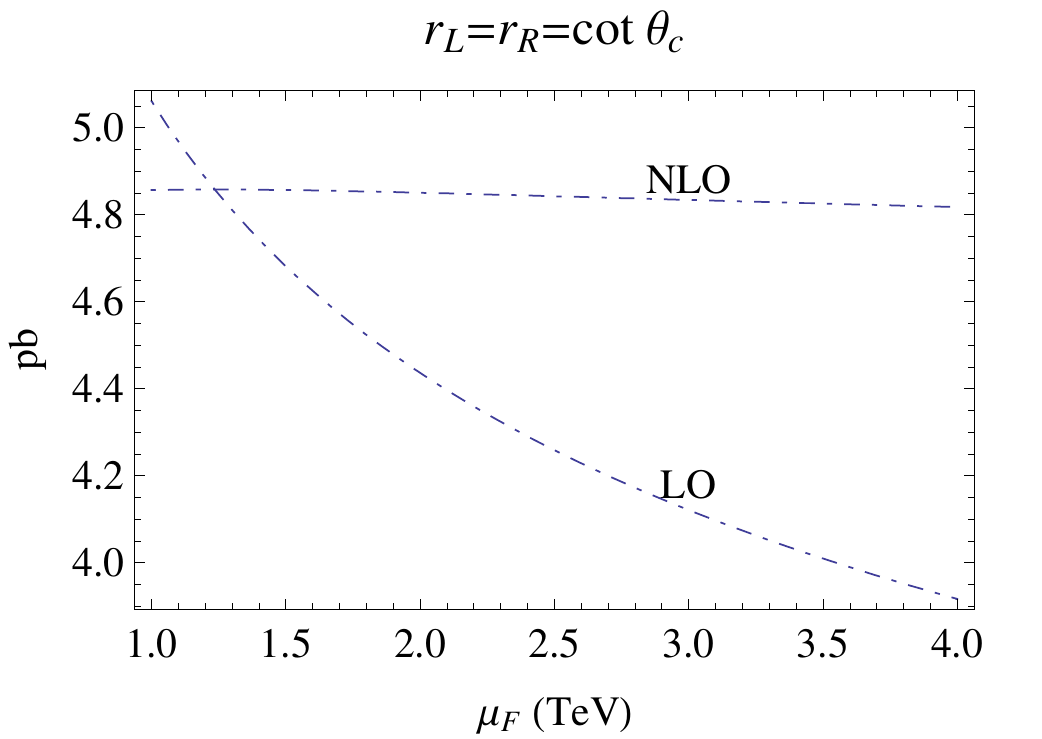}
\caption{Dependence of LO and NLO cross sections at the LHC ($\sqrt{s}=$ 7 TeV), as a function of factorization scale $\mu_F$
for $M_C$ = 2.0 TeV, $\sin^2\theta_c\vert_{\mu = 2.0\,{\rm TeV}} = 0.25$, and the three possible flavor-universal scenarios for the quark charge assignments. As expected, the NLO cross section has a much weaker (formally, two-loop) residual scale-dependence.}
\label{fig:scale-dependence}
\end{figure}

\begin{figure}
\includegraphics[width=2.3in]{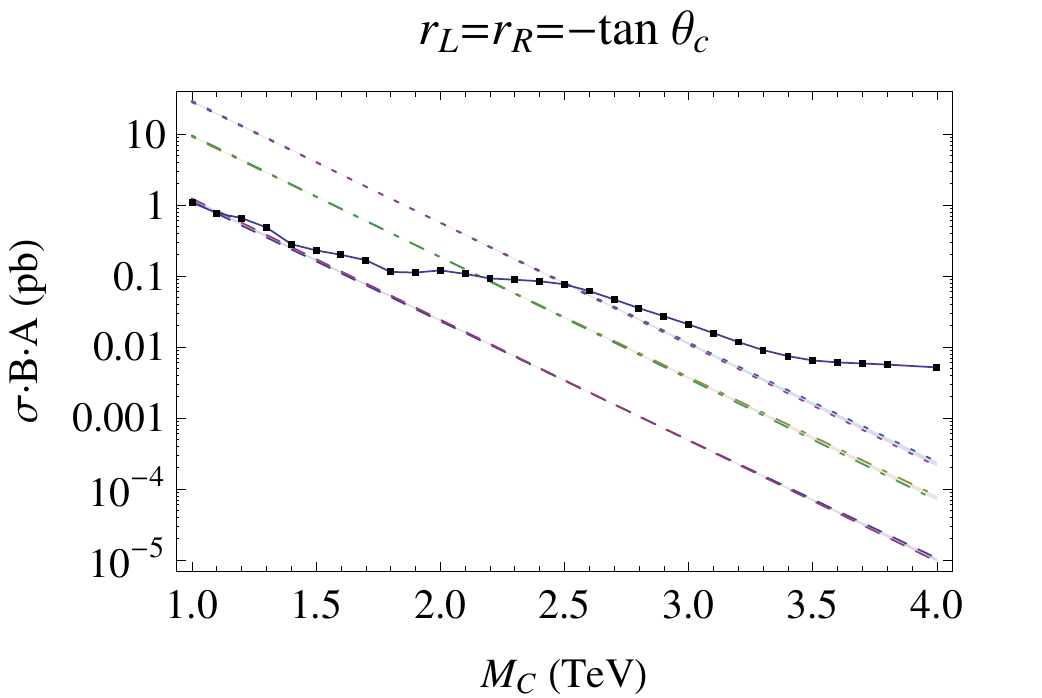}
\includegraphics[width=2.3in]{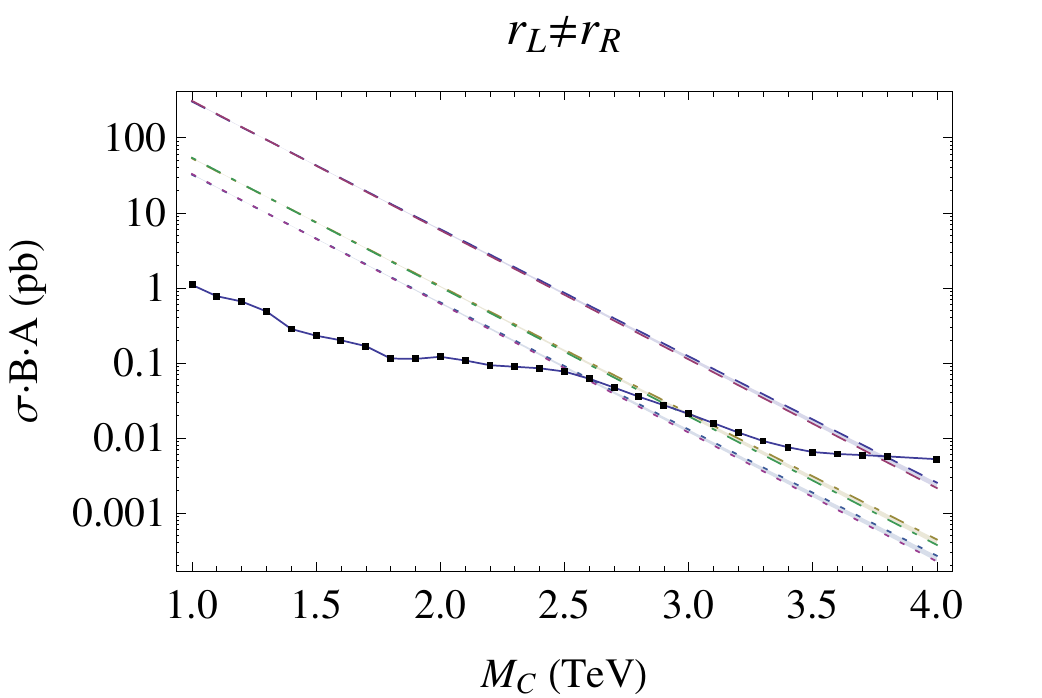}
\includegraphics[width=2.3in]{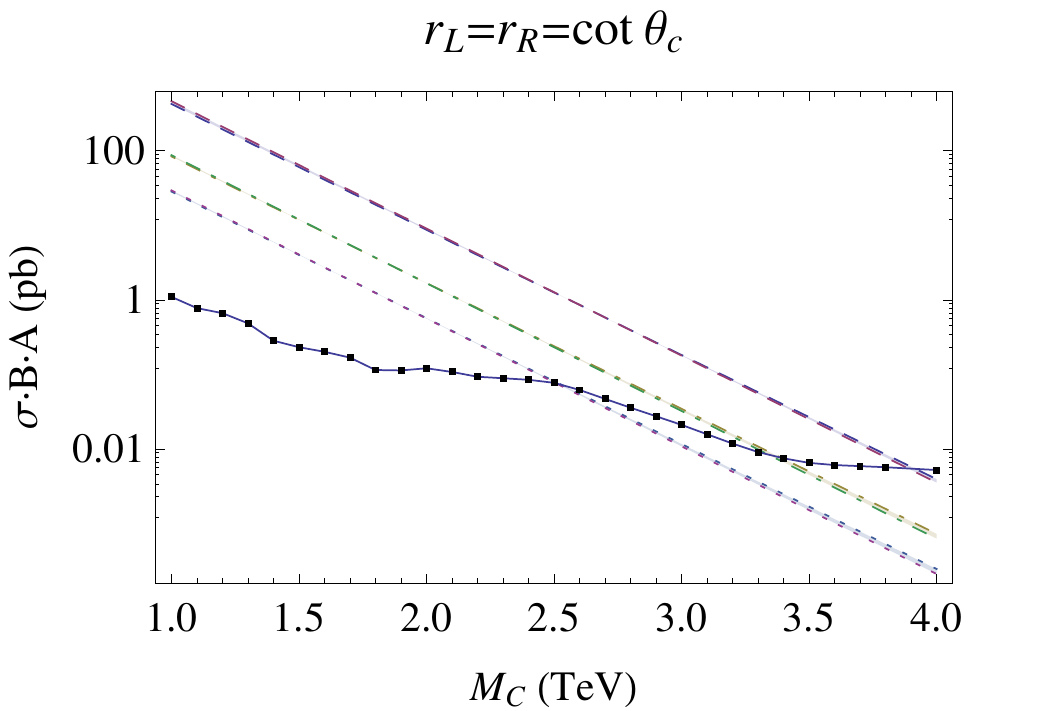}
\caption{NLO cross section times branching-ratio to quarks for on-shell coloron production at the LHC ($\sqrt{s}=$ 7 TeV), corrected for acceptance as described in the text. We consider the three possible flavor-universal scenarios for the quark charge assignments, take the renormalization scale $\mu$ to be equal to $M_C$, and plot $\sigma$ for $\sin^2\theta_c\vert_{\mu=M_C}=$ 0.05 (dashed), 0.25 (dot-dashed), and 0.5 (dotted). We plot these cross sections for $\mu_F$ ranging from $M_C/2$ to $2\,M_C$ and,  reflecting the weak dependence of the NLO cross section on the factorization scale, the resulting bands for each $\sin^2\theta_c$ are very narrow. To give a sense of current experimental reach, we plot the CMS \protect\cite{CMS} upper limit (solid line)
on the cross-section times dijet branching ratio for a narrow resonance. Note that the axigluon \protect\cite{Frampton:1987dn}
corresponds to the middle $r_L \neq r_R$ plot with $\sin^2\theta=0.5$ -- and hence a narrow axigluon resonance is
constrained to have a mass of order 2.6 TeV or higher.}
\label{fig:sigma}
\end{figure}

\begin{figure}
\includegraphics[width=2.3in]{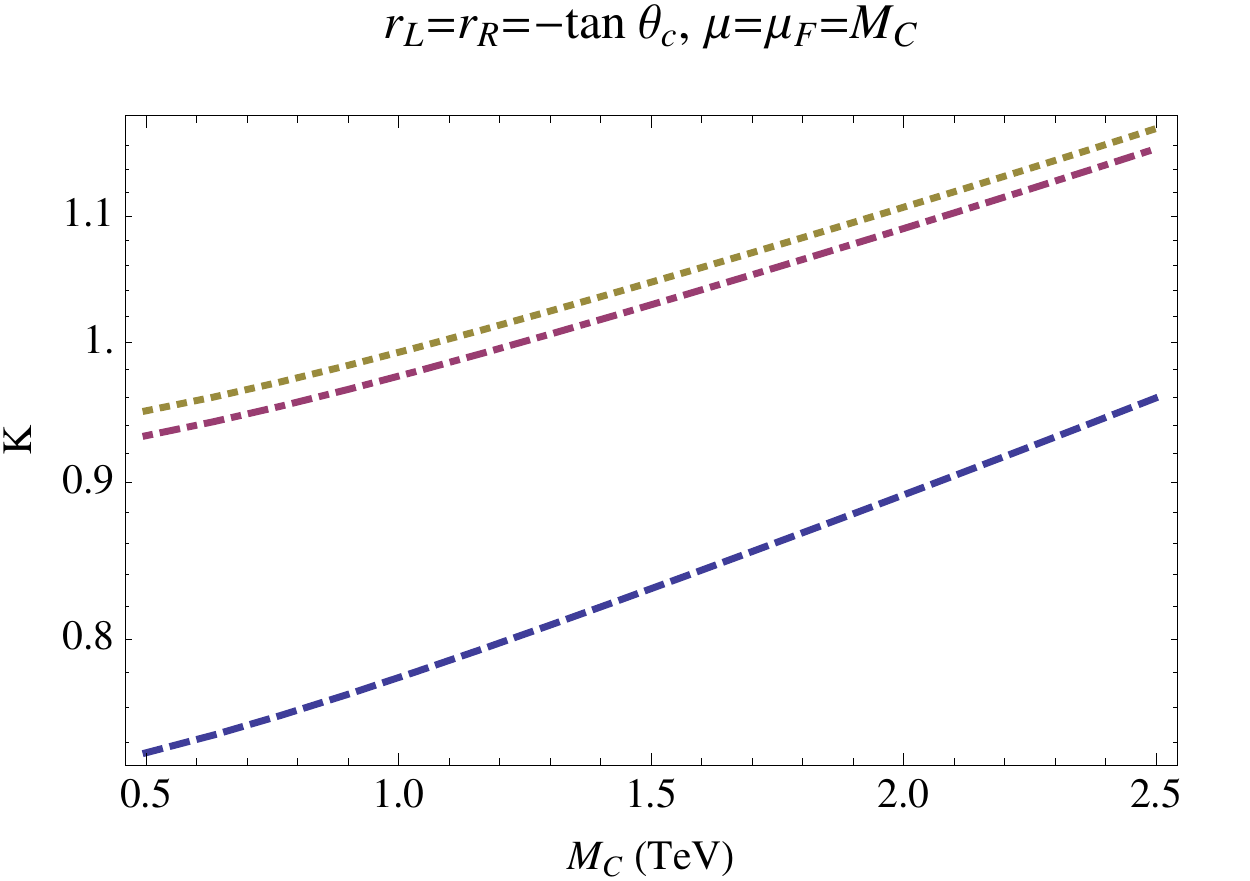}
\includegraphics[width=2.3in]{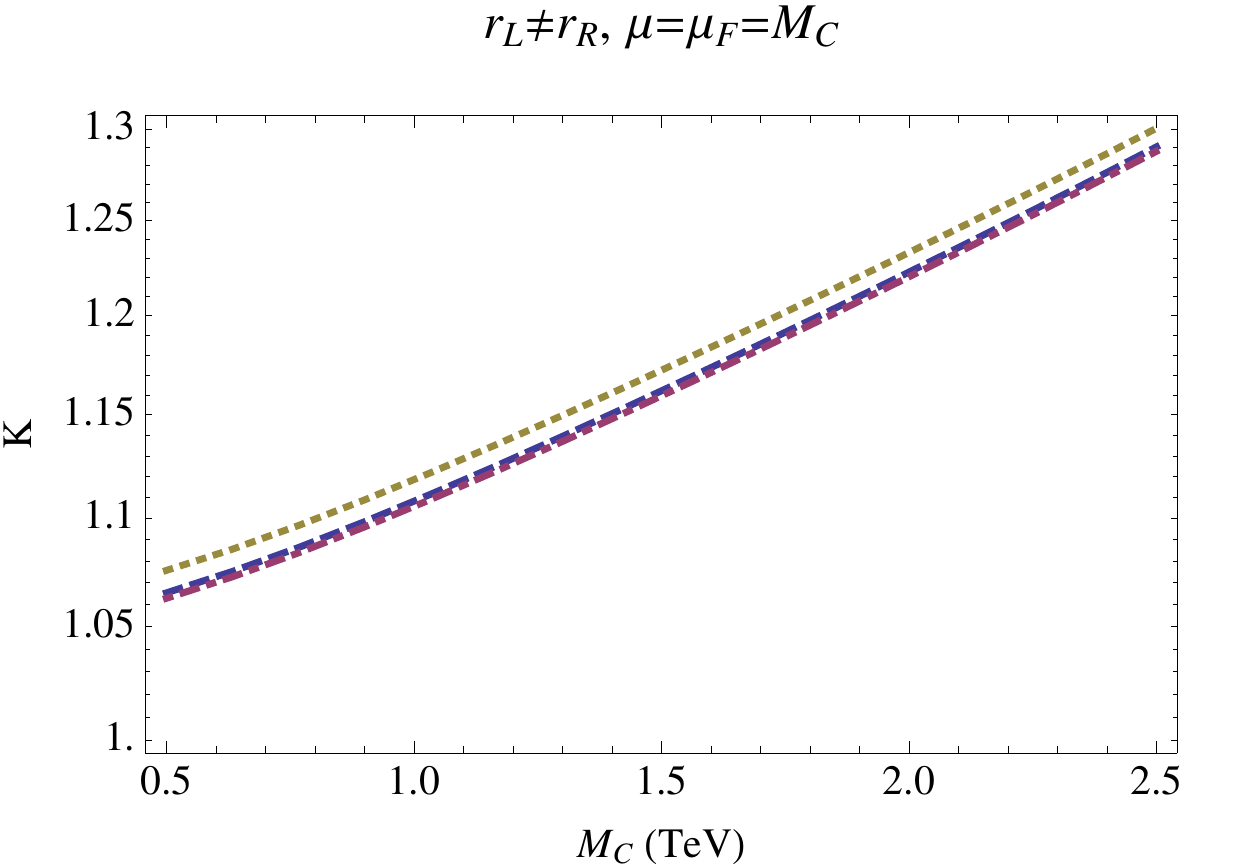}
\includegraphics[width=2.3in]{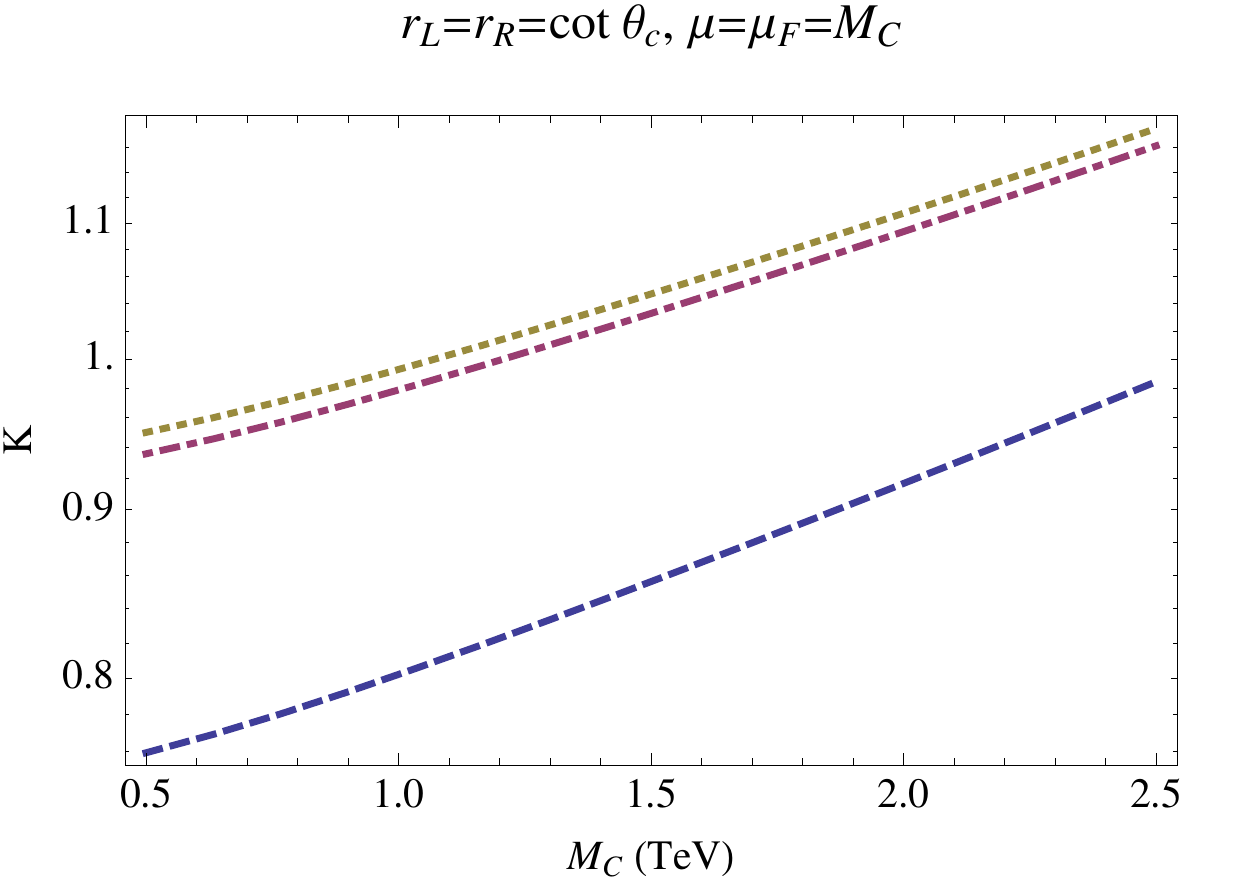}
\caption{``$K$-factor", the ratio of the NLO to LO cross section for coloron production at the LHC ($\sqrt{s}=7$ TeV), plotted as a function of $M_C$ for $\sin^2\theta_c$ = 0.05 (dashed), 0.25 (dot-dashed) and 0.50 (dotted), $\mu_F=M_C$, and the three different quark charge assignments.}
\label{fig:Kfactor}
\end{figure}

We now illustrate\footnote{For the purposes of illustration we use  the Mathematica package for CTEQ5 \protect\cite{Lai:1999wy} to evaluate the relevant parton distribution functions.} our results for the NLO coloron production cross-section in Figs.~\ref{fig:Q} - \ref{fig:Kfactor}.
In each figure we consider the three possible flavor-universal scenarios for quark charge assignment: $r_L=r_R=-\tan\theta_c$, $r_L\neq r_R$, and $r_L=r_R=\cot\theta_c$. All of the plots refer to coloron production at the LHC with $\sqrt{s}=7$ TeV. 

Notice that the perturbative expansion is only meaningful as long as $\sin\theta_c$ is neither too close to zero (where $g_{2s}\gg g_{1s}$) nor too close to one (where $g_{1s}\gg g_{2s}$). This is clear from Fig.~\ref{fig:Q}, in which we plot the quantity $Q$ defined in Eq.~(\ref{eq:QNLO}), for $\mu=M_C$: the contribution from the virtual corrections to the NLO cross section. The upper curve is for the $r_L\neq r_R$ scenario, whereas the almost identical lower curves are for $r_L=r_R=-\tan\theta_c$, and $r_L=r_R=\cot\theta_c$. For $\sin^2\theta_c\lesssim 0.05$ and $\sin^2\theta_c\gtrsim 0.95$ the virtual corrections become large, and the perturbative expansion in $\alpha_s$ breaks down. Since $\alpha_s\simeq 0.118$ at the $Z$ pole, these boundaries correspond to $g_{2s}\gtrsim 2.7$ and $g_{1s}\gtrsim 2.7$, respectively.


In Fig.~\ref{fig:scale-dependence}, we plot the $\mu_F$ dependence of the LO and NLO production cross sections
of a 2.0 TeV coloron (with $\sin^2\theta_c\vert_{\mu=2.1\, {\rm TeV}}$=0.25). The scale-dependence of the LO
cross section is of order 30\% while, as expected, the NLO cross section has a much weaker scale dependence, only of the order of 2\% percent.

In Fig.~\ref{fig:sigma} we plot the  cross section times branching ratio to quark jets as a function of $M_C$, allowing $\mu_F$ to vary from $M_C/2$ to $2M_C$.
Here, in order to compare to the experimental results of \cite{CMS} (shown as the solid line in the figures), we correct for the acceptance of the
detector by multiplying our partonic-level NLO production cross section by the factor
\begin{equation}
R = \frac{(\sigma(pp \to C)\cdot {\cal B} \cdot A)^{CMS}_{axigluon}}{\sigma^{LO}(pp\to C)_{axigluon}}~.
\end{equation}
In this expression,
$(\sigma(pp \to C)\cdot {\cal B} \cdot A)^{CMS}_{axigluon}$ is the CMS (LO) prediction for axigluon production cross section, times dijet branching ratio,
times acceptance\footnote{The CMS acceptance for isotropic decays is of order 0.6, independent of resonance mass \protect\cite{CMS}.} reported in \cite{CMS}, and $\sigma^{LO}(pp\to C)_{axigluon}$ is the  leading
order cross section in Eq. (\ref{CSfulldetect}) in the case of an axigluon ({\it i.e.} $r_L = -r_R = 1$), assuming the branching ratio to quarks ${\cal B}(C \to q\bar{q})=1$.\footnote{It is worth noting that there are examples of models with colorons which do not decay primarily to dijets, {\it e.g.}  \protect\cite{Kilic:2008pm}.}
The three sets of thin bands correspond to $\sin^2\theta_c\vert_{\mu=M_C}=$ 0.05 (dashed), 0.25 (dot-dashed), and 0.5 (dotted). Here, the weak residual $\mu_F$ dependence is shown by the narrowness of the bands. To give a sense of current experimental reach, we also show the 1 fb$^{-1}$ CMS upper bounds on the cross-section times di-jet branching ratio for a narrow resonance~\cite{CMS}. Note that the bound on the axigluon \cite{Frampton:1987dn}
corresponds to the $r_L \neq r_R$ plot with $\sin^2\theta=0.5$ -- and hence a narrow axigluon resonance is
constrained to have a mass of order 2.6 TeV or higher. The enhancement of the axigluon cross section at NLO
is responsible for the increase in the bound from of order 2.5 TeV as reported in \cite{CMS}.

Next, we compute the ``$K$-factor" for coloron production.
\begin{equation}
K(M_C,\sin\theta_c\vert_{\mu=M_C},\mu_F=M_C) \equiv \frac{\sigma^{NLO}(M_C,\sin\theta_c\vert_{\mu=M_C},\mu_F=M_C)}{\sigma^{LO}(M_C,\sin\theta_c\vert_{\mu=M_C},\mu_F=M_C)}~,
\label{eq:Kfactor}
\end{equation}
shown in Fig. \ref{fig:Kfactor} for $\sin^2\theta_c$ = 0.05 (dashed), 0.25 (dot-dashed) and 0.50 (dotted).
Again, we see that the NLO corrections are of order 30\%. In Appendix \ref{sec:Kfactors} we report the numerical
values of the $K$-factors corresponding to Fig. \ref{fig:Kfactor}, as well as those corresponding to the ATLAS
KK-gluon search reported in \cite{ATLAS-KK}.

\begin{figure}[h]
\begin{center}
\includegraphics[width=3.5in]{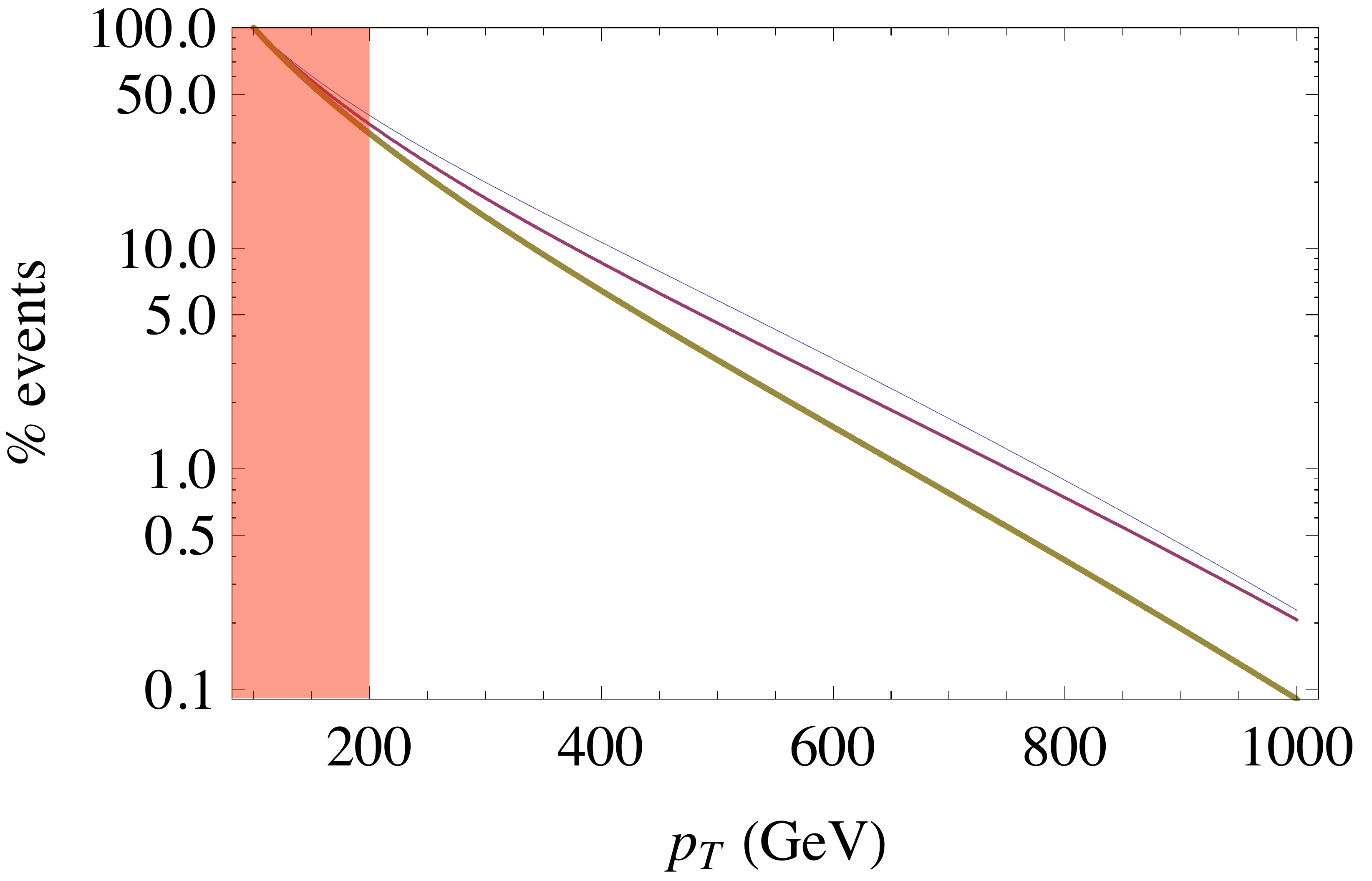}
\end{center}
\caption{Fraction of colorons produced with a $p_T$ greater than $p_{Tmin}$, as
a function of $p_{Tmin}$. The curves are for $M_C=1.2$ (highest, thin blue line), 2.0 (middle, medium purple line), and 3.0 TeV (lowest, thick green line), for the vectorial case $r_L=r_R=-\tan\theta_c$ and $\sin^2\theta_c = 0.05$. Note that of order 30\% of the colorons in
this mass range are produced with $p_T \ge 200$ GeV. As denoted by the red shaded region, below a $p_T$ of 200 GeV the corrections
become larger than 30\%, terms proportional to $\log(M^2_C/p^2_{Tmin})$
become large, and this fixed-order calculation becomes unreliable.
}
\label{fig:pTplotINT}.
\end{figure}

At leading order, the coloron is produced with zero transverse momentum. 
We may use our results to compute the $p_T$ spectrum in coloron production to leading
non-trivial order from Eq. (\ref{eq:dsigmadpt}).  Using these formulae, we may compute the {\it fraction} of colorons produced 
above a momentum
$p_{Tmin}$
\begin{equation}
{\cal P}(p_T \ge p_{Tmin}, M_C, \sin\theta_c\vert_{\mu=M_C},\mu_F=M_C) \equiv \frac{1}{\sigma^{NLO}(M_C,\sin \theta_c\vert_{\mu=M_C},\mu_F=M_C)} \int_{p_{Tmin}}^{p_{Tmax}} dp_T\, \frac{d\sigma}{dp_T}~,
\end{equation}
where $p_{Tmax}$ is the kinematic maximum transverse momentum (which depends on the coloron mass). 
For illustration, we plot this fraction for
vectorial colorons ($r_L=r_R=-\tan\theta_c$, with $\sin^2\theta_c=0.05$) with masses of 1.2,
2.0, and 3.0 TeV in Fig. \ref{fig:pTplotINT}. Note that of order 30\% of the colorons in
this model and mass range are produced with $p_T \ge 200$ GeV. Below a $p_T$ of 200 GeV the corrections
become larger than 30\%, terms proportional to $\log(M^2_C/p^2_{Tmin})$
become large, and this fixed-order calculation becomes unreliable.

In conclusion, we have reported the first complete calculation of QCD corrections to the production of a massive
color-octet vector boson.
Our next-to-leading-order calculation includes both virtual corrections as well as corrections arising from the emission of gluons and light quarks, and we have demonstrated the reduction in factorization-scale dependence relative to the leading-order
approximation used in previous hadron collider studies. In particular, we have shown that
the QCD NLO corrections to coloron production
are as large as 30\%, and that the residual factorization scale-dependence is reduced to of order 2\%.
We have also calculated the $K$-factor and the $p_T$ spectrum for coloron production, since these are
valuable for comparison with experiment.
Our computation applies
directly to the production of the massive color-octet vector bosons in axigluon, topcolor, and coloron models,
and approximately to the production of KK gluons in extra-dimensional models or colored technivector mesons in
technicolor models.
We look forward to future results from the LHC, and the possible discovery of colorons.

\section{Acknowledgments}

We thank Wayne W. Repko and Carl Schmidt for their input and discussions. A.F. also thanks
John M. Campbell for useful conversations during TASI 2011 summer school, where part of this work was completed. This work is supported  in part, by the US National Science Foundation under grant PHY-0854889.

\appendix
\label{FRsec}
\section{Feynman Rules\footnote{The Feynman rules discussed here are equivalent to those
in \protect\cite{Zhu:2011gd}, aside from those for the triple-coloron vertex which is not specified in
that reference.}} 
\label{sec:FRs}
\begin{figure}[h]
\begin{center}
\includegraphics[width=\textwidth]{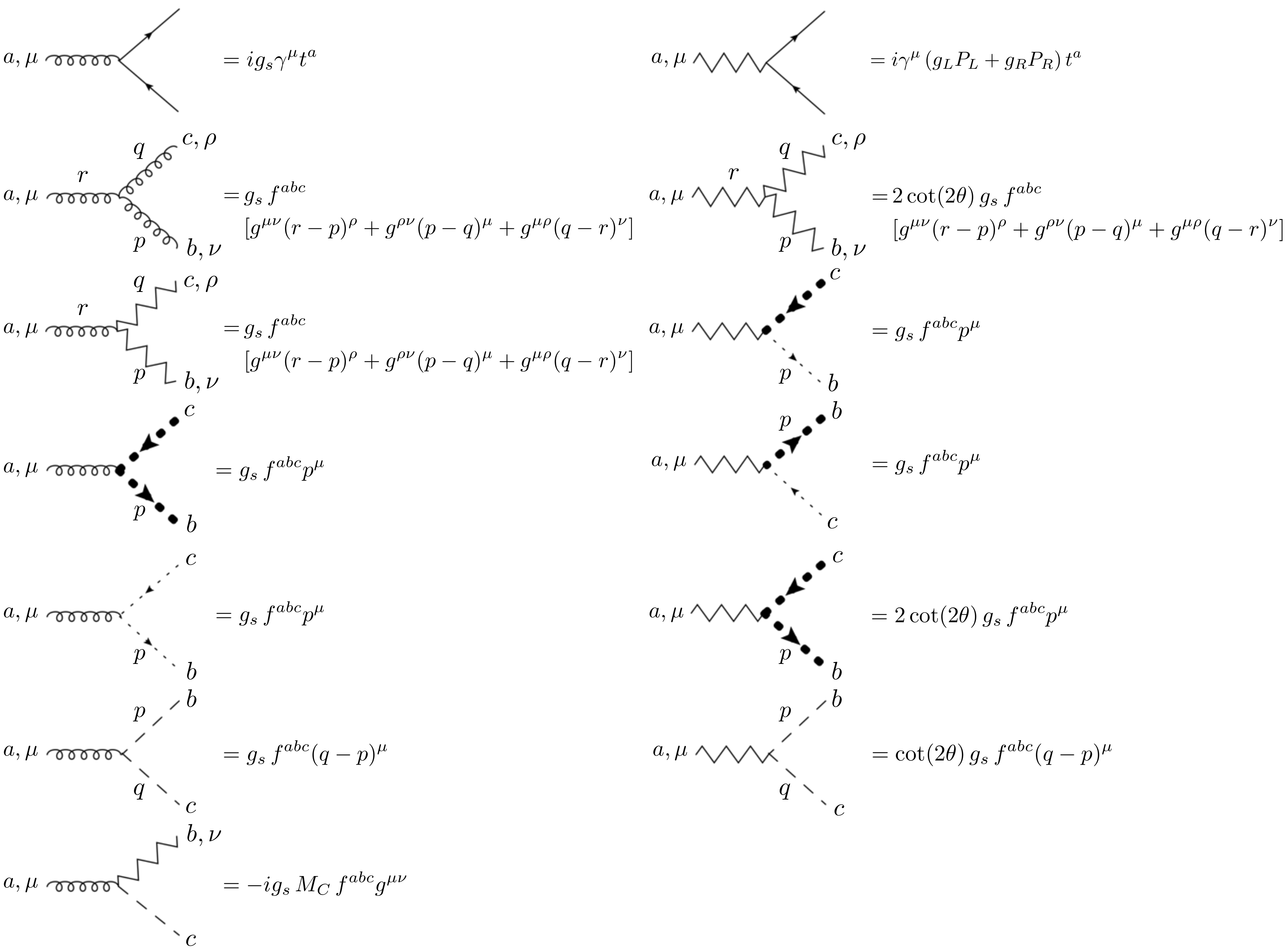}
\caption{Feynman rules for the trilinear vertices. In each diagram the momenta are toward the vertex. A gluon field is, as usual, represented by a coiling line; a coloron field is represented by a zigzag line. The coloron ghost is represented by a sequence of filled circles, and the eaten Goldstone bosons are represented by dashed lines.}
\label{fig:three}
\end{center}
\end{figure}

\begin{figure}[h]
\begin{center}
\includegraphics[width=\textwidth]{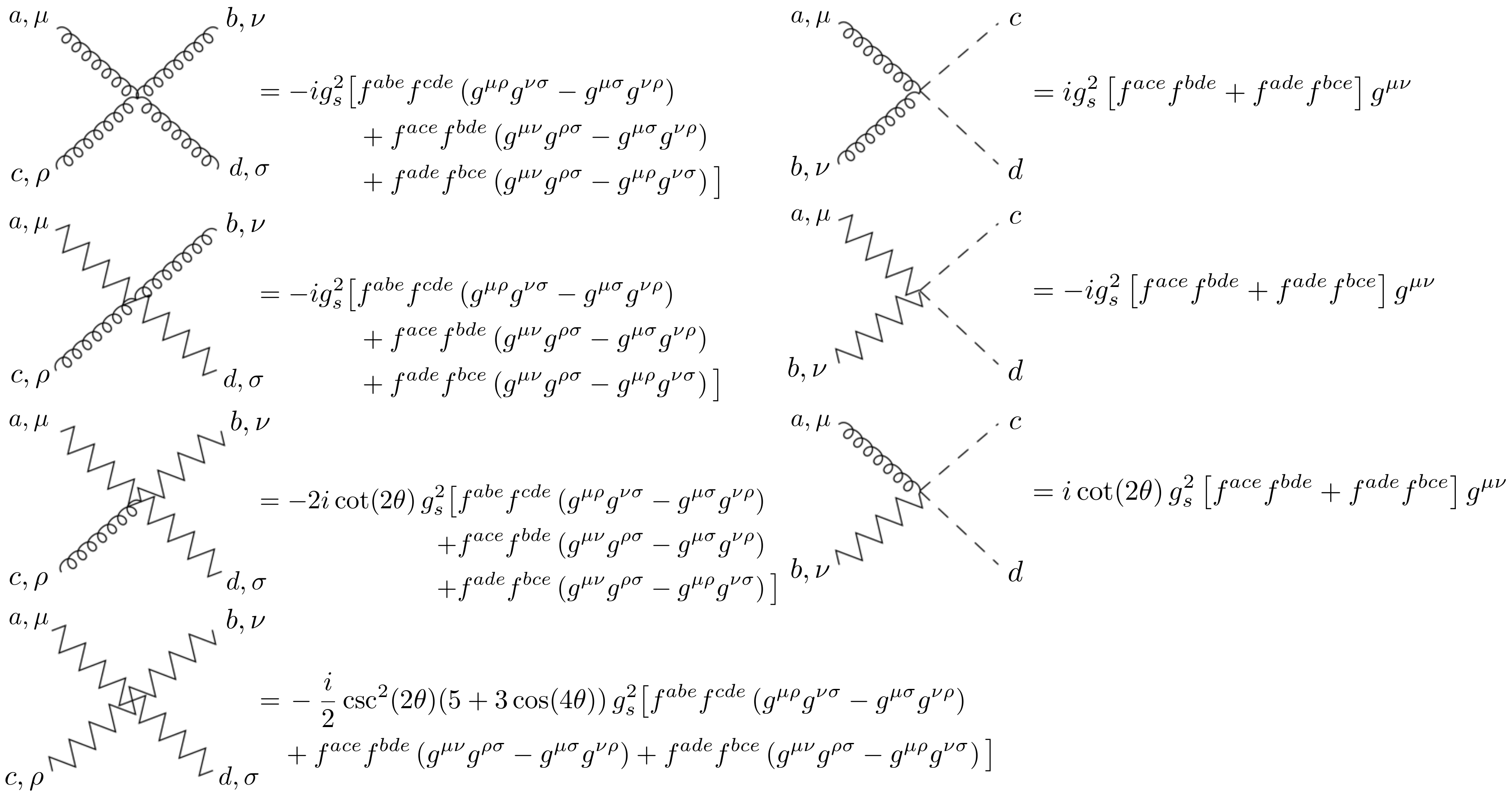}
\caption{Feynman rules for the quartic vertices. A gluon field is, as usual, represented by a coiling line; a coloron field is represented by a zigzag line.}
\label{fig:four}
\end{center}
\end{figure}

The Feynman rules for the trilinear and quartic vertices are shown in Figs.~\ref{fig:three} and~\ref{fig:four}, respectively. The coloron is represented by a zigzag line, the coloron ghost by a sequence of small circles, and the eaten Goldstone bosons by dashed lines.  All other particles are denoted as in QCD standard notation.

\section{Numerical Values of the $K$-Factor}
\label{sec:Kfactors}

\begin{figure}
\centering
\subfigure[\,$r_L=r_R=-\tan\theta_c$]{
\includegraphics[width=2in]{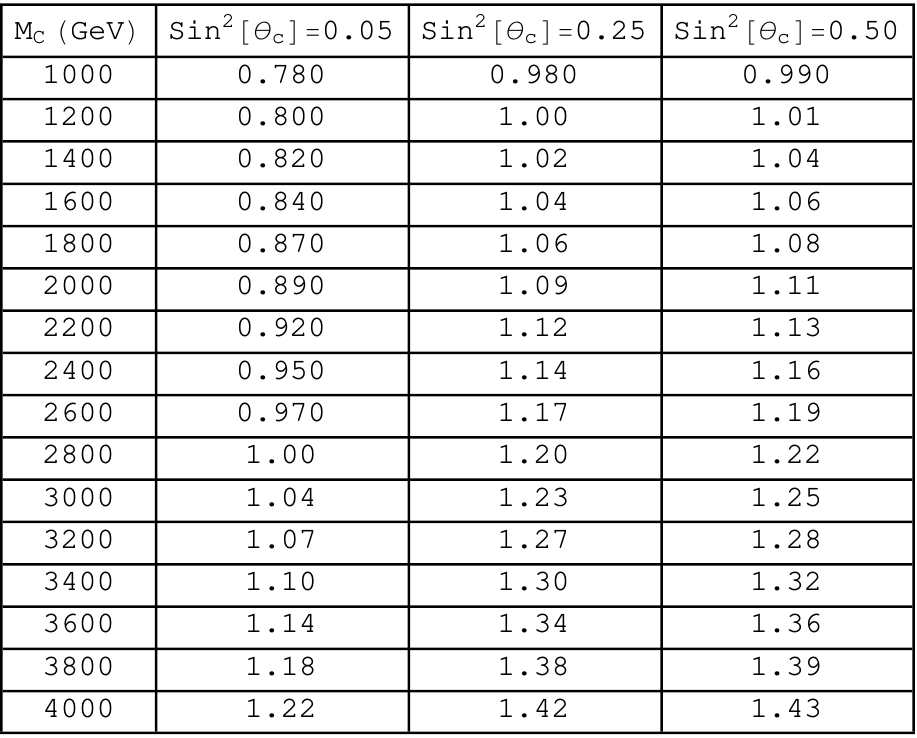}
\label{fig:K-factor-1}
}
\subfigure[\,$r_L\neq r_R$]{
\includegraphics[width=2in]{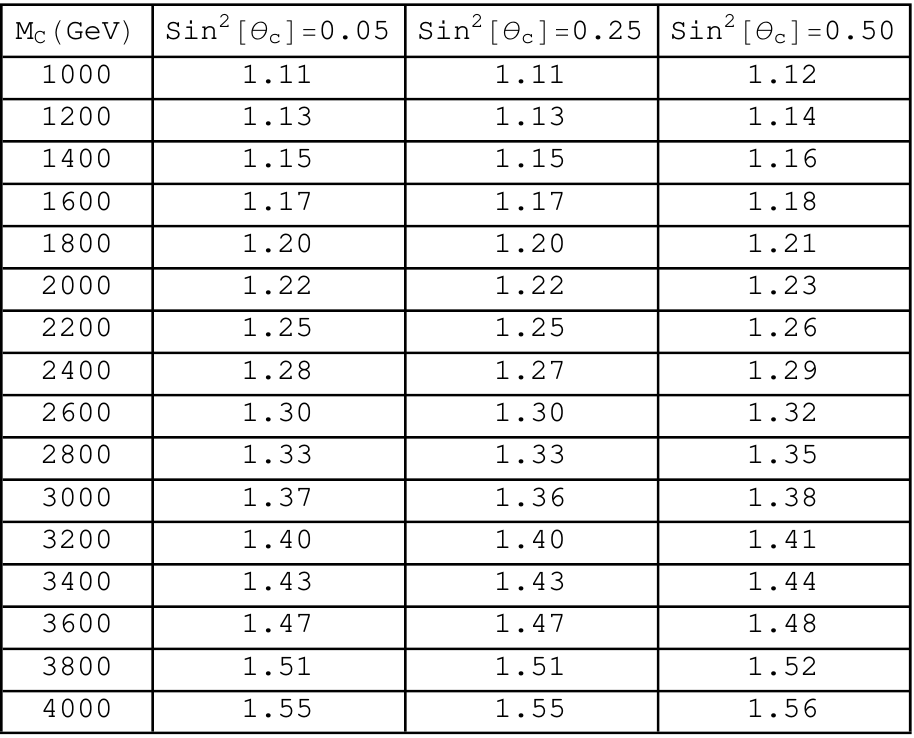}
\label{fig:K-factor-2}
}
\subfigure[\,$r_L= r_R=\cot\theta_c$]{
\includegraphics[width=2in]{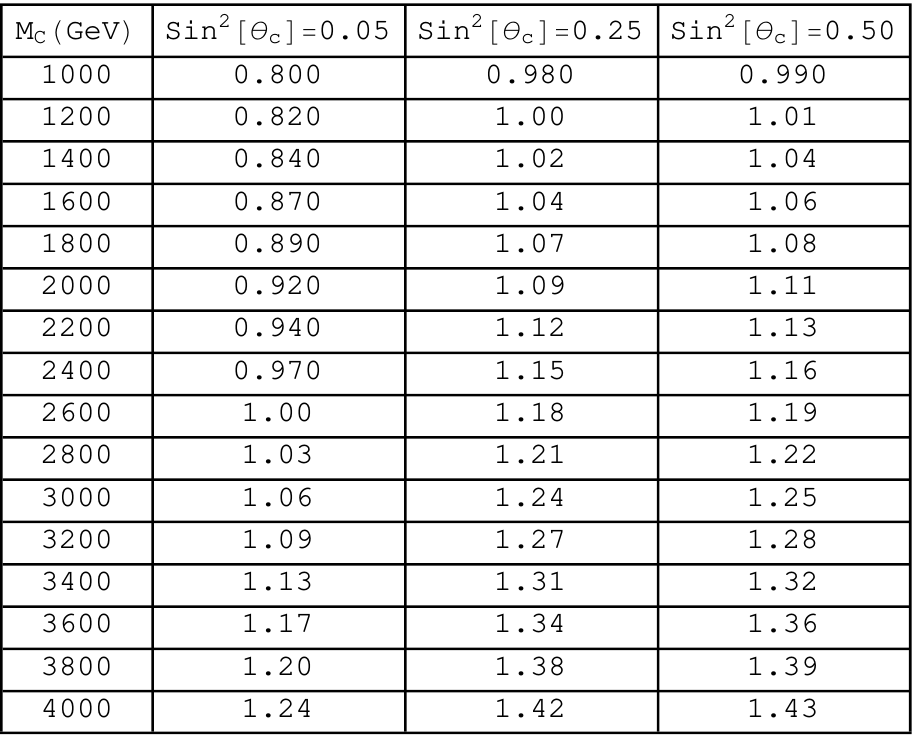}
\label{fig:K-factor-3}
}
\caption{$K$-factors for colorons of various masses and couplings. The classic ``axigluon"
\protect\cite{Frampton:1987dn} corresponds to $r_L \neq r_R$ and $\sin^2\theta_c=0.50$.}
\end{figure}

\begin{figure}
\begin{center}
\includegraphics[width=2in]{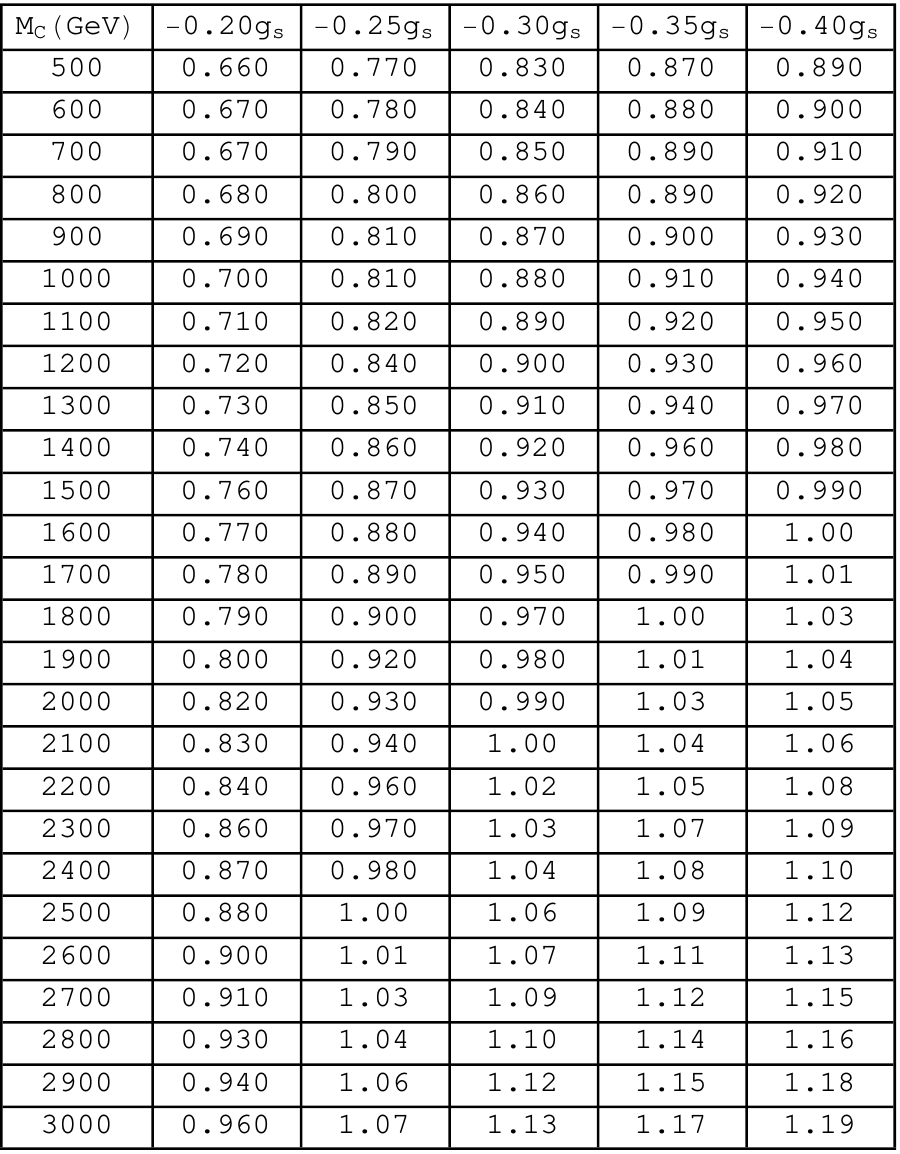}
\end{center}
\caption{$K$-factors for KK-gluons of various masses considered in \protect\cite{ATLAS-KK}. This calculation
is based on the theoretical framework of \protect\cite{Lillie:2007yh}, with the KK-gluon coupling
(specified in the column heading) varying between $-0.20 g_s$ and $-0.40 g_s$. }
\label{fig:KKgluon-K}.
\end{figure}

The numerical values of the $K$-factors for various values of the coloron mass and the three
patterns of coloron coupling are shown in Figs. \ref{fig:K-factor-1}, \ref{fig:K-factor-2}, and
\ref{fig:K-factor-3}. Finally, the values of the $K$-factor corresponding to the KK-gluons of \cite{Lillie:2007yh},
corresponding to the experimental search reported in \cite{ATLAS-KK}, are shown in Fig. \ref{fig:KKgluon-K}.

\end{document}